\documentclass[preprint,11pt,3p,A4paper]{elsarticle}

\usepackage[whole]{bxcjkjatype}
\usepackage{newtxtext}
\usepackage{amsmath,amssymb,bm,bbm}
\usepackage{mathtools}
\usepackage{booktabs}
\usepackage{lineno}
\usepackage{color}
\usepackage{subfigure}

\usepackage[ruled,vlined,linesnumbered]{algorithm2e}
\SetAlgoNoEnd
\DontPrintSemicolon

\biboptions{numbers,sort&compress}

\journal{Mechanical Systems and Signal Processing}

\begin{document}


\begin{frontmatter}

\title{{Hierarchical Bayesian model updating using Dirichlet process mixtures for structural damage localization}}

\author[label1]{Taro Yaoyama\corref{cor1}}\ead{yaoyama@g.ecc.u-tokyo.ac.jp}
\author[label1]{Tatsuya Itoi}
\author[label1]{Jun Iyama}

\cortext[cor1]{corresponding author}

\affiliation[label1]{
    organization={Graduate School of Engineering, The University of Tokyo},
    addressline={7-3-1, Hongo}, 
    city={Bunkyo-Ku},
    postcode={113-8656}, 
    state={Tokyo},
    country={Japan}
}

\begin{abstract}
Bayesian model updating provides a rigorous probabilistic framework for calibrating finite element (FE) models with quantified uncertainties, thereby enhancing damage assessment, response prediction, and performance evaluation of engineering structures.
Recent advances in hierarchical Bayesian model updating (HBMU) enable robust parameter estimation under ill-posed/ill-conditioned settings and in the presence of inherent variability in structural parameters due to environmental and operational conditions.
However, most HBMU approaches overlook multimodality in structural parameters that often arises when a structure experiences multiple damage states over its service life.
This paper presents a hierarchical Bayesian model updating (HBMU) framework that employs a Dirichlet process (DP) mixture prior on structural parameters (DP-HBMU).
DP mixtures are nonparametric Bayesian models that perform clustering without pre-specifying the number of clusters, incorporating damage state classification into FE model updating.
We formulate the DP-HBMU framework and devise a Metropolis-within-Gibbs sampler that draws samples from the posterior by embedding Metropolis updates for intractable conditionals due to the FE simulator.
The applicability of DP-HBMU to damage localization is demonstrated through both numerical and experimental examples.
We consider moment-resisting frame structures with beam-end fractures and apply the method to datasets spanning multiple damage states, from an intact state to moderate or severe damage state.
The clusters inferred by DP-HBMU align closely with the assumed or observed damage states.
The posterior distributions of stiffness parameters agree with ground truth values (in the numerical example) or observed fractures (in the experimental validation) while exhibiting substantially reduced uncertainty relative to a non-hierarchical baseline.
These results demonstrate the effectiveness of the proposed method in damage localization.
\end{abstract}


\end{frontmatter}

\section{Introduction} \label{sec:intro}

Finite element (FE) analysis plays a critical role in damage assessment, response prediction, and performance evaluation of engineering structures.
However, FE models generally exhibit some discrepancies relative to observed behavior of real-world structures, arising from modeling error, measurement noise, construction or manufacturing tolerances, inherent variability in material properties, variation in environmental or operational conditions, and structural deterioration or damage.
Bayesian model updating (BMU) addresses these discrepancies by calibrating structural parameters (e.g., stiffness, boundary conditions, mass) so that simulated responses align more closely with measurements, while quantifying uncertainties in parameters and predictions \cite{simoenE2015,huangY2019ase}.
BMU is generally formulated as follows.
Let $\mathbf{\Theta}$ denote a parameter set (including structural and statistical parameters), and let $\mathcal{X} = \{\mathbf{x}_1,...,\mathbf{x}_N\}$ be a set of observations.
Bayes' theorem yields the posterior distribution as $p(\mathbf{\Theta} \mid \mathcal{X}) \propto p(\mathcal{X} \mid \mathbf{\Theta}) ~ p(\mathbf{\Theta})$, where $p(\mathcal{X} \mid \mathbf{\Theta})$ is the likelihood, the probability of the observations under parameters $\mathbf{\Theta}$, and $p(\mathbf{\Theta})$ is a prior distribution that reflects domain knowledge about the system.
Given that the normalizing constant (evidence) $\int p(\mathcal{X} \mid \mathbf{\Theta}) ~ p(\mathbf{\Theta}) ~ \mathrm{d} \mathbf{\Theta}$ is typically intractable, the posterior is approximated using variational inference \cite{niP2021,hongF2025} or sampling methods such as Markov chain Monte Carlo (MCMC) \cite{beckJL2002} and sequential Monte Carlo (SMC) \cite{chingJ2007,carreraB2024}.

Recently, hierarchical Bayesian model updating (HBMU) has gained growing interest.
In typical settings, a prior is further characterized as $\mathbf{\Theta} \sim p(\mathbf{\Theta} \mid \mathbf{\Phi})$ with hyperparameters $\mathbf{\Phi} \sim p(\mathbf{\Phi})$, where $X \sim P$ denotes ``$X$ follows a distribution $P$''. 
One prominent application is sparse Bayesian learning (SBL) \cite{huangY2015,huangY2017a,huangY2017b}, which deals with ill-posed and ill-conditioned problems under data scarcity and measurement noise.
Huang and Beck~\cite{huangY2015} placed the automatic relevance prior on stiffness reduction parameters to induce their sparsity, enhancing damage assessment from noisy incomplete modal data.
HBMU has also been combined with multitask learning \cite{huangY2019cacie,xueS2023,yaoyamaT2024a}.
Huang et al.~\cite{huangY2019cacie} jointly addressed different measurements (tasks) by imposing a shared sparsity profile on stiffness reduction parameters, resulting in multitask SBL.
Yaoyama et al.~\cite{yaoyamaT2024a} presented a Bayesian multitask learning framework that partitions a whole structure into sub-structures (tasks) and jointly updates their parameters via a shared hierarchical prior to better exploit limited data.
Another line of research \cite{behmaneshI2015,songM2019,sedehiO2019,jiaX2022mssp,jiaX2023} addressed inherent variability in structural parameters due to environmental or experimental conditions (e.g., temperature or excitation amplitudes).
In these studies, each observation $\mathbf{x}_n$ was associated with individual structural parameters $\bm{\theta}_n$, and a shared prior $p(\bm{\theta} \mid \mathbf{\Phi})$ was placed over $\bm{\theta}_1, ..., \bm{\theta}_N$.
In most cases, $p(\bm{\theta} \mid \mathbf{\Phi})$ was taken to be a multivariate Gaussian and its mean and covariance were inferred via Gibbs sampling \cite{behmaneshI2015,songM2019}, asymptotic approximation \cite{sedehiO2019,jiaX2022mssp}, or variational inference \cite{jiaX2023}.

Previous HBMU studies, however, have rarely considered datasets that include multiple modes in structural parameters.
This limitation matters in structural health monitoring (SHM) applications because a real-world structure can experience multiple damage states during its service life, ranging from an intact state to moderate or severe damage states. 
Dealing with such multimodality requires a more flexible prior than a single-mode Gaussian and motivates a hierarchical Bayesian approach that separates damage states in the parameter space while quantifying and reducing uncertainty within each state.

This paper presents an HBMU framework that employs a Dirichlet process mixture model (DPMM) as a prior on structural parameters.
The DPMM is a nonparametric Bayesian clustering model in which the number of clusters is not fixed a priori but is inferred from the observations \cite{escobarMD1995,nealRM2000,tehYW2010}.
The DPMM prior can cluster structural parameters according to possible damage states while enhancing uncertainty-aware parameter estimation in FE models.
In the SHM literature, Mei et al.~\cite{meiLF2025} recently proposed a damage detection approach that applies DPMM‑based clustering directly to observed transmissibility functions.
In contrast, the present study infers unobserved structural parameters and constructs the DPMM in the parameter space.
This formulation enables joint inference of cluster assignments and model parameters, allowing the associated uncertainties to be propagated in a coherent manner.
Clustering in the parameter space then allows knowledge transfer across multiple observations within the same damage state, which can lead to reduced posterior uncertainty.
In the following, we formulate the HBMU framework based on DPMM (termed DP-HBMU) and devise an efficient Gibbs sampling procedure.
We further examine its applicability to structural damage localization, especially in the settings of stress resultant-based model updating \cite{yaoyamaT2024a,yaoyamaT2024b,yaoyamaT2024isrerm}, through both numerical and experimental examples.

The remainder of the paper is organized as follows.
Section 2 introduces the proposed DP-HBMU together with its Gibbs sampling algorithm.
Section 3 presents an illustrative example of damage localization in a planar moment frame, comparing against a non-hierarchical baseline.
Section 4 applies the DP-HBMU to experimental (dynamic) data from a one-bay, two-story steel frame, demonstrating accurate localization of beam-end fractures and estimation of rotational stiffnesses with reduced, well-quantified uncertainty. Section 5 concludes and outlines directions for future work.

\section{Proposed Framework}\label{sec:method}

\subsection{Problem formulation of hierarchical Bayesian model updating}

Consider an engineering system parameterized by system parameters (e.g., stiffness), $\bm{\theta} \in \mathbb{R}^D$.
The response of the system, $\mathbf{x}$, is assumed to be represented by a simulator (e.g., an FE solver), $h(\bm{\theta})$.
Let $\mathcal{X} = \{\mathbf{x}_n\}_{n=1}^{N} \subset \mathbb{R}^M$ denote a set of observed responses from the real system and $\bm{\theta}_n$ denote its associated (underlying) parameter.
The generative process of $\{\mathbf{x}_n\}$ is then described as
\begin{align}
    \mathbf{x}_n = h(\bm{\theta}_n) + \mathbf{e}_n, \quad n = 1, ..., N,
    \label{eq:gen}
\end{align}
where $\mathbf{e}_n$ denotes residuals including measurement and modeling error.
We assume that $\mathbf{e}_n$ independently and identically follows a zero-mean Gaussian distribution, $\mathbf{e}_n \sim \mathcal{N}(\cdot \mid \mathbf{0}, \beta^{-1}\mathbf{I})$, where $\beta$ is a precision parameter and $\mathbf{I}$ denotes an equivalent matrix.
Eq.~(\ref{eq:gen}) enables us to consider the stochasticity of $\bm{\theta}_n$ across observations, which originates from varying environmental (or experimental) conditions.
In the typical HBMU framework, we introduce a common prior as $\bm{\theta}_n \sim p(\cdot \mid \bm{\Phi})$, where $\mathbf{\Phi}$ denotes a set of parameters, resulting in the following generative model,
\begin{equation}
    \begin{aligned}
        \mathbf{x}_n &~ \sim ~ \mathcal{N}(\cdot \mid h(\bm{\theta}_n), \beta^{-1} \mathbf{I}), \quad n = 1,...,N, \\
        \bm{\theta}_n &~ \sim ~ p(\cdot \mid \mathbf{\Phi}), \quad n = 1,...,N. \\
    \end{aligned}
    \label{eq:gen2}
\end{equation}
To infer the whole set of parameters $\{\{\bm{\theta}_n\}, \mathbf{\Phi}, \beta\}$, we need to specify a hierarchical prior on $\mathbf{\Phi}$.
In the following, we assume that this prior follows a Dirichlet process and formulate a DP-HBMU framework.

\begin{figure}[!t]
    \centering
    \includegraphics[width=0.25\linewidth]{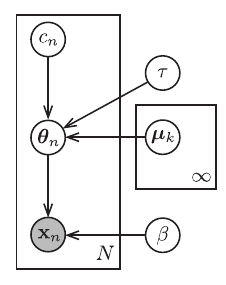}
    \caption{Graphical model representing a generative model in the DP-HBMU framework.}
    \label{fig:graphical}
\end{figure}

\subsection{Dirichlet process mixtures for hierarchical Bayesian model updating}

We consider that the stochastic model $p(\cdot \mid \mathbf{\Phi})$ exhibits multiple modes across different observations (in the SHM context, multiple damage states).
To address this, one may adopt a mixture model for $p(\cdot \mid \mathbf{\Phi})$, such as Gaussian mixtures.
Typical mixture models, however, often require a prior specification of the number of classes (modes), $K$, which is not straightforward in most engineering applications.
Determining $K$ via model selection criteria (e.g., BIC) is a possible approach; however, it generally requires repeating the inference procedure for multiple candidate values of $K$, which can be computationally prohibitive in simulation-based model updating problems.

We therefore introduce a Dirichlet process mixture model (DPMM) for the prior over $\{\bm{\theta}_n\}$ with the following form \cite{nealRM2000},
\begin{equation}
    \begin{aligned}
        \bm{\theta}_n ~ & \sim ~ p(\cdot \mid \bm{\phi}_n), \quad n = 1, ..., N, \\
        \bm{\phi}_n ~ & \sim ~ G(\cdot), \quad n = 1, ..., N, \\
        G ~ & \sim ~ \mathrm{DP}(\alpha, G_0),
    \end{aligned}
    \label{eq:gen-dpmm}
\end{equation}
where $\bm{\phi}_n \in \mathbf{\Phi}$ denotes a parameter vector, and $G$ denotes a random probability measure (distribution) over the space of $\bm{\phi}_n$.
$\mathrm{DP}(\alpha, G_0)$ denotes a Dirichlet process (DP) over $G$ with a scale parameter $\alpha ~ (>0)$ and a base measure $G_0$.
The DP randomly generates distributions themselves, and is thus called a ``distribution over distributions'' \cite{nealRM2000}.
The expectation of $G(\cdot)$ according to $\mathrm{DP}(\alpha, G_0)$ equals $G_0(\cdot)$, and the variance of $G(\cdot)$ is controlled by $\alpha$---if $\alpha$ is smaller, the DP has much more probability mass concentrated on the proximity of $G_0$.

A key property of the Dirichlet process is that any draw $G \sim \mathrm{DP}(\alpha, G_0)$ is almost surely discrete \cite{blackwellD1973}; that is, it places its probability mass on a countable set of atoms in the $\bm{\phi}$-space.
This suggests that two different observations $\mathbf{x}_i$ and $\mathbf{x}_j$ can share the same parameters; that is, $\bm{\phi}_i = \bm{\phi}_j$, which reflects the clustering behavior induced on a set of $\{\bm{\phi}_n\}$.
Let $\{\widehat{\bm{\phi}}_k\}_{k=1}^{K}$ denote a set of distinct elements (atoms) in $\{\bm{\phi}_n\}_{n=1}^N$, and $\{c_n : c_n \in \{1, ..., K\}, ~ n = 1, ..., N\}$ denote a set of class labels assigned to $\{\mathbf{x}_n\}_{n=1}^N$.
Clearly, $\bm{\phi}_n = \widehat{\bm{\phi}}_{c_n}$.
Marginalizing over $G$ in the generative model (\ref{eq:gen-dpmm}), we have the following equivalent form,
\begin{equation}
    \begin{aligned}
        \bm{\theta}_n ~ & \sim ~ p(\cdot \mid \widehat{\bm{\phi}}_{c_n}), \quad n = 1, ..., N \\
        c_1, ..., c_N ~ & \sim ~ \mathrm{CRP}(\alpha), \\
        \widehat{\bm{\phi}}_k ~ & \sim ~ G_0(\cdot), \quad k = 1, ..., \infty.
    \end{aligned}
    \label{eq:gen-crp}
\end{equation}
where $\mathrm{CRP}(\alpha)$ denotes the Chinese restaurant process (CRP) with a scale parameter $\alpha$.
Sampling $c_n$ from $\mathrm{CRP}(\alpha)$ works according to
\begin{align}
    P(c_n = k \mid c_1, ..., c_{n-1}, c_{n+1}, ..., c_N) =
    \begin{dcases}
        \frac{N_k^{\backslash n}}{n - 1 + \alpha} & \text{if }~ k \in \{c_j\}_{j \neq n} \\
        \frac{\alpha }{n - 1 + \alpha} & \text{if }~ k \notin \{c_j\}_{j \neq n}
    \end{dcases},
    \label{eq:classlabel}
\end{align}
where $N_k^{\backslash n} := \sum_{j \neq n} \delta(c_j = k)$ with $\delta(\cdot)$ denoting the indicator function that equals 1 when its argument is true and 0 otherwise.
It assigns the observation to an existing class with probability proportional to the number of other observations already in that class, or to a new class with probability proportional to $\alpha$.
The CRP prior on the labels $\{c_n\}$ eliminates the need to pre-define the number of classes, leading to \textit{infinite} mixtures \cite{tehYW2010}.
Since $\alpha$ is critical in the clustering behavior, we place the Gamma prior $p(\alpha) = \mathrm{Ga}(\alpha \mid a_\alpha, b_\alpha)$ and infer it jointly in the posterior inference.

In this study, we assume a Gaussian distribution for the conditional of $\bm{\theta}_n$ as $p(\bm{\theta}_n \mid \bm{\mu}_{c_n}, \tau) = \mathcal{N}(\bm{\theta}_n \mid \bm{\mu}_{c_n}, \tau^{-1} \mathbf{I})$, where $\bm{\mu}_k ~ (= \widehat{\bm{\phi}}_k)$ denotes the mean vector for the $k$th cluster, and $\tau$ denotes a precision parameter assumed to be shared across all classes $k = 1, ..., \infty$.
For $\{\bm{\mu}_k\}$ and $\tau$, we set a normal-gamma distribution as a conjugate prior,
\begin{align}
    p(\bm{\mu}_k, \tau) = p(\bm{\mu}_k \mid \tau) ~ p(\tau)    
    = \mathcal{N}(\bm{\mu}_k \mid \bm{\mu}_0, (\rho\tau)^{-1} \mathbf{I}) ~ \mathrm{Ga}(\tau \mid a_\tau, b_\tau),
    \label{eq:normal-gamma}
\end{align}
where $\mathrm{Ga}(\cdot \mid a, b)$ denotes a Gamma distribution with a shape parameter $a$ and rate parameter $b$.
The conditional $\mathcal{N}(\bm{\mu}_k \mid \bm{\mu}_0, (\rho\tau)^{-1})$ corresponds to $G_0$ in the model (\ref{eq:gen-crp}).
We also set a conjugate prior for $\beta$ as $p(\beta) = \mathrm{Ga}(\beta \mid a_\beta, b_\beta)$.
These configurations result in a Dirichlet process Gaussian mixture model (DPGMM) with the following form,
\begin{equation}
    \label{eq:gen-dphbmu}
    \begin{aligned}
        \mathbf{x}_n ~ & \sim ~ \mathcal{N}(\cdot \mid h(\bm{\theta}_n), \beta^{-1} \mathbf{I}), \quad n = 1, ..., N, \\
        \bm{\theta}_n ~ & \sim ~ \mathcal{N}(\cdot \mid \bm{\mu}_{c_n}, \tau^{-1} \mathbf{I}), \quad n = 1, ..., N,\\
        c_1, ..., c_N ~ & \sim ~ \mathrm{CRP}(\alpha), \\
        \bm{\mu}_k ~ & \sim ~ \mathcal{N}(\cdot \mid \bm{\mu}_0, (\rho\tau)^{-1}\mathbf{I}), \quad k = 1 , ..., \infty, \\
        \tau ~ & \sim ~ \mathrm{Ga}(\cdot \mid a_\tau, b_\tau), \\
        \beta ~ & \sim ~ \mathrm{Ga}(\cdot \mid a_\beta, b_\beta), \\
        \alpha ~ & \sim ~ \mathrm{Ga}(\cdot \mid a_\alpha, b_\alpha).
    \end{aligned}
\end{equation}
The corresponding graphical model is illustrated in Figure~\ref{fig:graphical}.

\subsection{Metropolis-within-Gibbs sampler for DP-HBMU}

\begin{algorithm}[!t]
    \caption{Metropolis-within-Gibbs sampler for DP-HBMU} \label{alg:dp-hbmu}
    \KwIn{Data $\{\boldsymbol{x}_n\}_{n=1}^N$; hyperparameters $(\rho, \bm{\mu}_0, a_\beta, b_\beta, a_\tau, b_\tau, a_\alpha, b_\alpha)$.}
    \KwOut{Posterior samples $\{\mathbf{\Theta}^{(t)}\}_{t=1}^T \sim p(\mathbf{\Theta} \mid \{\mathbf{x}_n\}_{n=1}^N)$}
    \For{$t=1$ \KwTo $T$}{
        \tcp{\footnotesize Update cluster assignments via the split--merge algorithm (see \ref{sec:app})}
        Randomly select indices $i$ and $j$ from $\{1, ..., N\}$ \\
        \If{$c^{(t-1)}_i = c^{(t-1)}_j$}{
            Construct split proposal via restricted Gibbs sampling: $\{c^\ast_n\} \gets \{c^\mathrm{split}_n\}$\\
            Compute the proposal probability $q(\{c^\mathrm{split}_n\} \mid \{c^{(t-1)}_n\})$ \\
            Set $q(\{c^{(t-1)}_n\} \mid \{c^\mathrm{split}_n\}) = 1$
        }
        \Else{
            Construct merge proposal $\{c^\ast_n\} \gets \{c^\mathrm{merge}_n\}$ \\
            Compute the (reverse) proposal probability $q(\{c^{(t-1)}_n\} \mid \{c^\mathrm{merge}_n\})$ \\
            Set $q(\{c^\mathrm{merge}_n\} \mid \{c^{(t-1)}_n\}) = 1$ \\
        }
        Accept $\{c^{(t)}_n\} \gets \{c^\ast_n\}$ according to the probability $\zeta$ \tcp*{Eq.(\ref{eq:split-merge_ratio_0})}
        $K \gets$ the number of distinct elements in $\{c^{(t)}_n\}$ \\
        Sample $\alpha^{(t)} ~ \sim ~ p(\cdot \mid \{c^{(t)}_n\})$ \tcp*{Eq.(\ref{eq:kappa}--\ref{eq:alpha})}
        \For{$k=1$ \KwTo K}{
            Sample $\bm{\mu}_k^{(t)} \sim p(\cdot \mid \{\bm{\theta}^{(t-1)}_n\}, \{c_n^{(t)}\}, \tau^{(t-1)})$ \tcp*{Eq.(\ref{eq:sample_mu_k})}
        }
        Sample $\tau^{(t)} \sim p(\cdot \mid \{\bm{\theta}^{(t-1)}_n\}, \{c_n^{(t)}\})$ \tcp*{Eq.(\ref{eq:sample_tau})}
        \For{$n=1$ \KwTo N}{
            Run pCN to obtain $\bm{\theta}^{(t)}_n \sim p(\cdot \mid \mathbf{x}_n, c_n^{(t)}, \{\bm{\mu}_k^{(t)}\}, \tau^{(t)}, \beta^{(t-1)})$ \tcp*{Eqs.(\ref{eq:sample_theta_n}--\ref{eq:pcn_ratio})} 
        }
        Sample $\beta^{(t)} \sim p(\cdot \mid \{\mathbf{x}_n\}, \{\bm{\theta}_n^{(t)}\})$ \tcp*{Eq.(\ref{eq:sample_beta})}
    }
\end{algorithm}

Let a complete set of parameters in the model (\ref{eq:gen-dphbmu}) denoted by $\mathbf{\Theta} := \{\{\bm{\theta}_n\}, \{c_n\}, \{\bm{\mu}_k\}, \tau, \beta, \alpha\}$.
To infer the posterior $p(\mathbf{\Theta \mid \mathcal{X}})$, we adopt a class of MCMC algorithms, Gibbs sampling.
Gibbs sampling iteratively draws samples for each variable from its conditional probability given other variables, and is known to be robust to high dimensionality in the parameters.
This algorithm is efficient especially when every conditional distribution is available in a closed form and can be sampled directly.
When one or more conditionals lack a closed-form expression, the sampler can still be implemented by embedding other MCMC updates such as the Metropolis-Hastings algorithm, resulting in a ``Metropolis-within-Gibbs'' scheme.

The Metropolis-within-Gibbs sampler for $p(\mathbf{\Theta} \mid \mathcal{X})$ proceeds as follows.
\begin{enumerate}
    \item
    \textit{Sampling $c_n$}.
    Let $K$ denote the current number of clusters, and the class labels are assumed to be relabeled such that $c_n \in \{1, ..., K\}$.
    According to Jain and Neal \cite{jainS2004}, the updates to class labels $\{c_n\}$ follow a \textit{restricted split–merge} Metropolis–Hastings scheme.
    At each update, two distinct observations $i$ and $j$ are randomly selected from $\{1, ..., N\}$, and only the observations currently belonging to the same clusters as $i$ and $j$,
    \begin{align}
        \mathcal{S} = \{n : n \in \{1,...,N\}, n \neq i, n \neq j, c_n\in \{c_i, c_j\} \},
        \label{eq:S}
    \end{align}
    are considered.
    If $i$ and $j$ belong to the same cluster, i.e., $c_i = c_j = k$, we construct a split proposal $\{c^\mathrm{split}_n\}$ by \textit{restricted} Gibbs sampling, wherein $c^\mathrm{split}_i = k$, $c^\mathrm{split}_j = K + 1$ (a new class), and $c^\mathrm{split}_n = c_n$ for $n \notin \mathcal{S} \cup \{i, j\}$ are fixed; and $c^\mathrm{split}_n$ for $n \in \mathcal{S}$ is sampled from $\{k, K + 1\}$.
    If $i$ and $j$ belong to different clusters, i.e., $c_i \neq c_j$, we construct a merge proposal $\{c^\mathrm{merge}_n\}$, in which $c^\mathrm{merge}_n = c_j$ for $n \in \mathcal{S} \cup \{i, j\}$ and $c^\mathrm{merge}_n = c_n$ for $n \notin \mathcal{S} \cup \{i, j\}$.
    We then evaluate the proposal $\{c^\ast_n\}$ (either $\{c^\mathrm{split}_n\}$ or $\{c^\mathrm{merge}_n\}$) using the Metropolis--Hastings acceptance probability given by
    \begin{align}
        \zeta = \min \left\{1,
        \frac{q(\{c_n\} \mid \{c^\ast_n\})}{q(\{c^\ast_n\} \mid \{c_n\})}
        \frac{p(\{\bm{\theta}_n\} \mid \{c^\ast_n\}, \tau)}{p(\{\bm{\theta}_n\} \mid \{c_n\}, \tau)}
        \frac{P(\{c^\ast_n\} \mid \alpha)}{P(\{c_n\} \mid \alpha)}
        \right\},
        \label{eq:split-merge_ratio_0}
    \end{align}
    where $P(\{c_n\} \mid \alpha)$ denotes the CRP prior; $p(\{\bm{\theta}_n\} \mid \{c_n\}, \tau)$ denotes the (marginal) likelihood; and $q(\{c^\ast_n\} \mid \{c_n\})$ denotes the proposal probability.
    The readers are referred to \ref{sec:app} for details.
    %
    \item
    \textit{Sampling} $\alpha$.
    The sampling procedure for the CRP parameter $\alpha$ follows Escobar and West \cite{escobarMD1995}.
    \begin{align}
        \kappa ~ &\sim ~ \mathcal{B}(\cdot \mid \alpha + 1, N) \label{eq:kappa} \\
        \alpha ~ &\sim ~ \pi_\kappa \mathrm{Ga}(\cdot \mid a_\alpha + K, b_\alpha - \log \kappa) +
        (1 - \pi_\kappa) \, \mathrm{Ga}(\cdot \mid a_\alpha + K - 1, b_\alpha - \log \kappa) \label{eq:alpha}
    \end{align}
    where $\kappa$ is an auxiliary variable; 
    $\mathcal{B}(\cdot \mid a, b)$ denotes a beta distribution with shape parameters $(a, b)$;
    and $\pi_\kappa = (a_\alpha + K - 1) / (N(b_\alpha - \log \kappa) + a_\alpha + K - 1)$ is a mixing weight of two Gamma distributions.
    The readers interested in details are referred to Escobar and West \cite{escobarMD1995}.
    %
    \item \textit{Sampling} $\bm{\mu}_k$.
    Due to the conjugate priors in Eq.~(\ref{eq:normal-gamma}), $\bm{\mu}_k$ can directly be sampled from a normal distribution as follows.
    \begin{align}
        p(\bm{\mu}_k \mid \{\bm{\theta}_n\}, \{c_n\}, \tau) &=
        \mathcal{N} \left(
            \bm{\mu}_k
        ~ \middle| ~
            \frac{N_k}{N_k + \rho} \overline{\bm{\theta}}_k + \frac{\rho}{N_k + \rho} \bm{\mu}_0, ~ \frac{1}{\tau(N_k + \rho)} \mathbf{I}
        \right), \label{eq:sample_mu_k}
    \end{align}
    where $N_k := \sum_{n=1}^N \delta(c_n = k)$ and $\overline{\bm{\theta}}_k := N_k^{-1}\sum_{n=1}^N \delta(c_n = k) ~ \bm{\theta}_n$.
    %
    \item \textit{Sampling} $\tau$.
    Similarly to $\bm{\mu}_k$, $\tau$ can directly be sampled from a gamma distribution as
    \begin{align}
        ~&~p(\tau \mid \{\bm{\theta}_n\}, \{c_n\}) \nonumber \\
        =&~
        \mathrm{Ga} \left( \tau ~ \middle| ~
            a_\tau + \frac{ND}{2}, ~
            b_\tau + \frac{1}{2} \sum_{k = 1}^K 
            \left\{
                \sum_{n = 1}^N \delta(c_n = k) \|\bm{\theta}_n - \overline{\bm{\theta}}_k\|^2 +
                \frac{\rho N_k}{\rho + N_k} \|\overline{\bm{\theta}}_k - \bm{\mu}_0\|^2
            \right\}
        \right). \label{eq:sample_tau}
    \end{align}
    %
    \item \textit{Sampling} $\bm{\theta}_n$.
    The conditional probability of $\bm{\theta}_n$ can be described as 
    \begin{align}
        p(\bm{\theta}_n \mid \mathbf{x}_n, c_n, \{\bm{\mu}_k\}, \tau, \beta)
        ~\propto& ~~
        p(\mathbf{x}_n \mid \bm{\theta}_n, \beta) ~ p(\bm{\theta}_n \mid \bm{\mu}_{c_n}, \tau) \nonumber \\
        ~=& ~~
        \mathcal{N}(\mathbf{x}_n \mid h(\bm{\theta}_n), \beta^{-1} \mathbf{I}) ~ \mathcal{N}(\bm{\theta}_n \mid \bm{\mu}_{c_n}, \tau^{-1} \mathbf{I}).
        \label{eq:sample_theta_n}
    \end{align}
    Since the conditional depends on the simulator $h(\cdot)$, a closed-form solution is generally unavailable.
    We thus update $\bm{\theta}_n$ by embedding a Metropolis step within the Gibbs sampling procedure.
    Specifically, we employ the preconditioned Crank-Nicolson (pCN) proposal \cite{cotterSL2013,carreraB2024}.
    It is applicable to sampling from the posterior when the prior is Gaussian, and known to remain effective in the high-dimensional parameter space.
    In this study, the pCN proposal of a candidate $\bm{\theta}^\ast_n$ is described as
    \begin{align}
        q(\bm{\theta}^\ast_n \mid \bm{\theta}_n) =
        \mathcal{N}(\bm{\theta}_n^\ast \mid \bm{\mu}_{c_n} + \sqrt{1 - s^2} (\bm{\theta}_n - \bm{\mu}_{c_n}), ~ s^2\tau^{-1}\mathbf{I}),
    \end{align}
    where $s \in (0, 1]$ denotes a step size. This proposal ensures the following equality,
    \begin{align}
        q(\bm{\theta}^\ast_n \mid \bm{\theta}_n) ~ p(\bm{\theta}_n \mid \bm{\mu}_{c_n}, \tau) =  q(\bm{\theta}_n \mid \bm{\theta}^\ast_n) ~ p(\bm{\theta}^\ast_n \mid \bm{\mu}_{c_n}, \tau).
    \end{align}
    The left-hand and right-hand sides thus cancel out in the numerator and denominator of the acceptance probability in the Metropolis update.
    The acceptance probability reduces to
    \begin{align}
        \eta = \min \left\{1, ~ \frac{\mathcal{N}(\mathbf{x}_n \mid h(\bm{\theta}^\ast_n), \beta^{-1} \mathbf{I})}{\mathcal{N}(\mathbf{x}_n \mid h(\bm{\theta}_n), \beta^{-1} \mathbf{I})} \right\}.
        \label{eq:pcn_ratio}
    \end{align}
    %
    \item \textit{Sampling} $\beta$.
    Due to the conjugate prior, $\beta$ can directly be sampled from a Gamma distribution as
    \begin{align}
        p(\beta \mid \{\mathbf{x}_n\}, \{\bm{\theta}_n\}) =
        \mathrm{Ga} \left( \beta
        ~ \middle| ~
            a_\beta + \frac{NM}{2},
            b_\beta + \frac{1}{2} \sum_{n = 1}^N \|\mathbf{x}_n - h(\bm{\theta}_n)\|^2
        \right).
        \label{eq:sample_beta}
    \end{align}
\end{enumerate}

The above procedure is summarized in Algorithm~\ref{alg:dp-hbmu}.

\subsection{Remarks}

\subsubsection{Adaptation of the step size $s$ in the pCN proposal}

In this study, the step size $s$ of the pCN algorithm is adaptively tuned during Gibbs sampling procedure to achieve a target acceptance rate $\lambda^\ast$ according to the following rule \cite{andrieuC2008},
\begin{align}
    \mathrm{logit}(s^{(t+1)}) = \mathrm{logit}(s^{(t)}) + (\lambda^{(t)} - \lambda^\ast) \, t^{-v},
    \label{eq:update_s}
\end{align}
where $\mathrm{logit}(s) = \log(s/(1-s))$ denotes the logit function, which ensures that $s$ remains within the interval $(0, 1]$.
Here, $\lambda^{(t)}$ denotes the mean acceptance rate computed from samples $\{\bm{\theta}_n^{(t)}\}$ over the most recent $T_\mathrm{ref}$ iterations (in this study, $T_\mathrm{ref} = 50$), and $v$ is a diminishing updating rate.
This adaptation is performed only during $t \in (T_\mathrm{ref}, T_\mathrm{burn}]$ with $T_\mathrm{burn}$ denoting the burn-in period, after which $s$ is fixed.
Following Andrieu and Thoms \cite{andrieuC2008}, the diminishing rate $v$ is selected from the plausible interval $(0.5, 1.0]$; in this study, we set $v = 0.6$.
Throughout this paper, the target acceptance rate is set to $\lambda^\ast = 0.8$.
Although this value is higher than those commonly used in the literature (e.g., \cite{cotterSL2013}), it reflects the hierarchical structure of the proposed model; that is, the prior $\mathcal{N}(\cdot \mid \bm{\mu}_{c_n}, \tau^{-1}\mathbf{I})$ is adaptively centered around plausible values of $\bm{\theta}_n$, which tends to yield higher acceptance probabilities.

\subsubsection{Relabeling}\label{sec:relabeling}

In the inference of mixture models, we sometimes suffer from the label-switching problem; that is, the same partition state can allow multiple equivalent ways of labeling, causing labels to switch during MCMC iterations.
For example, the partition $\{\{c_1, c_2\}, \{c_3\}, \{c_4\}\}$ is represented in multiple ways, e.g., $\{1, 1, 2, 3\}$ or $\{2, 2, 1, 3\}$.
We address this issue by employing the relabeling algorithm proposed by Stephens \cite{stephensM2000} as follows.

\textit{Relabeling algorithm.}
We focus on the most probable number of clusters, denoted by $\widehat{K}$, and retain posterior samples of $c_n$ and $\bm{\mu}_k$ conditional on $K=\widehat{K}$, denoted by $\{\{c_n^{(t)}\}_{n=1}^N\}_{t=1}^{T'}$ and $\{\{\bm{\mu}_k^{(t)}\}_{k=1}^{\widehat{K}}\}_{t=1}^{T'}$.
Let $\nu_t$ denote a permutation of the cluster label set $\{1,...,\widehat{K}\}$ at iteration $t$, which maps $k$ to a relabeled assignment $k' = \nu_t(k) \in \{1,...,\widehat{K}\}$.
For each iteration $t$, we define an $N \times \widehat{K}$ assignment matrix under a permutation $\nu_t$, $\mathbf{P}(t;\nu_t)$, whose $(n,k)$-th element is given by
\begin{align}
p_{nk}(t;\nu_t) = \delta (\nu_t(c_n^{(t)})=k).
\end{align}
Starting from initial permutations $\{\nu_t\}_{t=1}^{T'}$ (e.g., identity permutations), repeat the following procedure until convergence.
\begin{enumerate}
    \item Find a probability matrix $\mathbf{Q}$ that minimizes the sum of Kullback--Leibler divergences,
    \begin{align}
        \mathbf{Q}^\star
        = \arg\min_{\mathbf{Q}} \sum_{t=1}^{T'} \mathrm{KL} \left(\mathbf{P}(t;\nu_t) \| \mathbf{Q}\right) = \arg\min_{\mathbf{Q}} \sum_{t=1}^{T'} \sum_{n=1}^N \sum_{k=1}^{\widehat{K}} p_{nk}(t;\nu_t)\log\frac{p_{nk}(t;\nu_t)}{q_{nk}},
    \end{align}
    where each row of $\mathbf{Q}$ sums to one, i.e., $\sum_k q_{nk} = 1$, with $q_{nk}$ denoting the $(n, k)$-th element in $\mathbf{Q}$.
    This minimization results in
    \begin{align}
        q_{nk} = \frac{1}{T'}\sum_{t=1}^{T'} p_{nk}(t;\nu_t).
    \end{align}
    \item For each $t = 1,..., T'$, update the permutation $\nu_t$ so as to minimize the Kullback--Leibler divergence between $\mathbf{Q}$ and $\mathbf{P}(t; \nu_t)$,
    \begin{align}
        \mathrm{KL}(\mathbf{P}(t; \nu_t) \| \mathbf{Q})
        = \sum_{n=1}^N \sum_{k=1}^{\widehat{K}} p_{nk}(t; \nu_t) \log \frac{p_{nk}(t; \nu_t)}{q_{nk}} = -\sum_{n=1}^N \delta(\nu_t(c_n^{(t)}) = k) \, \log q_{nk}.
    \end{align}
\end{enumerate}

\subsubsection{Modeling of within-cluster dispersion}

Since the present study focuses on structural damage detection, the number of observations associated with each cluster (damage state) is generally limited.
In the following examples, each cluster contains 2--8 observations.
This makes it challenging to estimate cluster-specific dispersion parameters like $\tau_k^{-1} \mathbf{I}$ or $\mathbf{\Sigma}_k$.
Therefore, we adopt a simplified model in which all clusters share a common precision parameter $\tau$, assuming an isotropic covariance structure.
However, this modeling may introduce potential limitations.
In situations where the true parameter variability greatly differs across clusters, distinct clusters may be spuriously merged, or a single cluster may be unnecessary split.
Moreover, the within-cluster variability may be overestimated or underestimated.
Such clustering issues can bias the estimation of model parameters $\bm{\theta}_n$ or introduce misspecified uncertainty quantification.

As future works, we will examine extensions that allow for cluster-specific isotropic covariances $\tau^{-1}_k$, as well as more refined models in which the full covariance matrix $\mathbf{\Sigma}_k$ are inferred using Wishart priors. We note that the effectiveness of these extensions may depend on the amount of data available within each cluster.

\section{Illustrative Example}\label{sec:num}

\subsection{Problem setup: stress resultant-based model updating}

In this section, we demonstrate the DP-HBMU in the context of \textit{stress resultant--based} Bayesian model updating \cite{yaoyamaT2024a,yaoyamaT2024b,yaoyamaT2024isrerm}.
This example is intended as a representative demonstration of the proposed framework, and its applicability to other structural systems or damage mechanisms depends on problem-specific situations such as model identifiability.
The problem setup is described as follows.
We assume that both accelerometers and dynamic strain gauges are installed on the target structure.
System identification techniques are applied jointly to acceleration and strain measurements to extract modal properties in terms of displacements and stress resultants.
These quantities are termed \textit{modal displacement} (MD) and \textit{modal stress-resultant} (MSR), respectively.
In this study, we concentrate on the bending-moment component of the stress resultant and call the corresponding mode shapes the \textit{modal bending moment} (MBM).
The MBM is computed from elastic section modulus and Young's modulus, both assumed to be known a priori (see subsection \ref{sec:exp-target} for computational details).
Based on the MD and MBM, we define \textit{normalized} MBM (nMBM) as
\begin{align}
    \bar{\mathbf{r}} = \mathbf{r} / \|\mathbf{d}\|
    \label{eq:nmbm}
\end{align}
where $\|\cdot\|$ denotes the $L_2$ norm of a vector.
The nMBM---also referred to as \textit{local stiffness}~\cite{iyamaJ2021ES,iyamaJ2021JCSHM}---represents the relationship between elemental stress and nodal displacement and therefore serves as a highly sensitive indicator of elemental stiffness, enhancing damage localization \cite{yaoyamaT2024a,yaoyamaT2024b}.

Stress resultant--based model updating exploits nMBM measurements to infer elemental stiffness parameters. 
In this study, we consider multiple observations $\mathcal{X} := \{\overline{\mathbf{r}}_n\}_{n=1}^N$ with the corresponding simulator $h: \mathbf{\bm{\theta}}_n \mapsto \overline{\mathbf{r}}_n$, and aim to infer the rotational stiffness of several beam ends for each observation, $\bm{\theta}_n$.
Note that the stress resultant--based framework does not require observations of $\overline{\mathbf{r}}_n$ at all beam or column ends. The impact of sensor locations and incomplete measurements on estimation uncertainty has been investigated in Yaoyama et al.~\cite{yaoyamaT2024isrerm}.

\subsection{Target structure and data}

\begin{figure}[!t]
    \centering
    \includegraphics[width=0.50\linewidth]{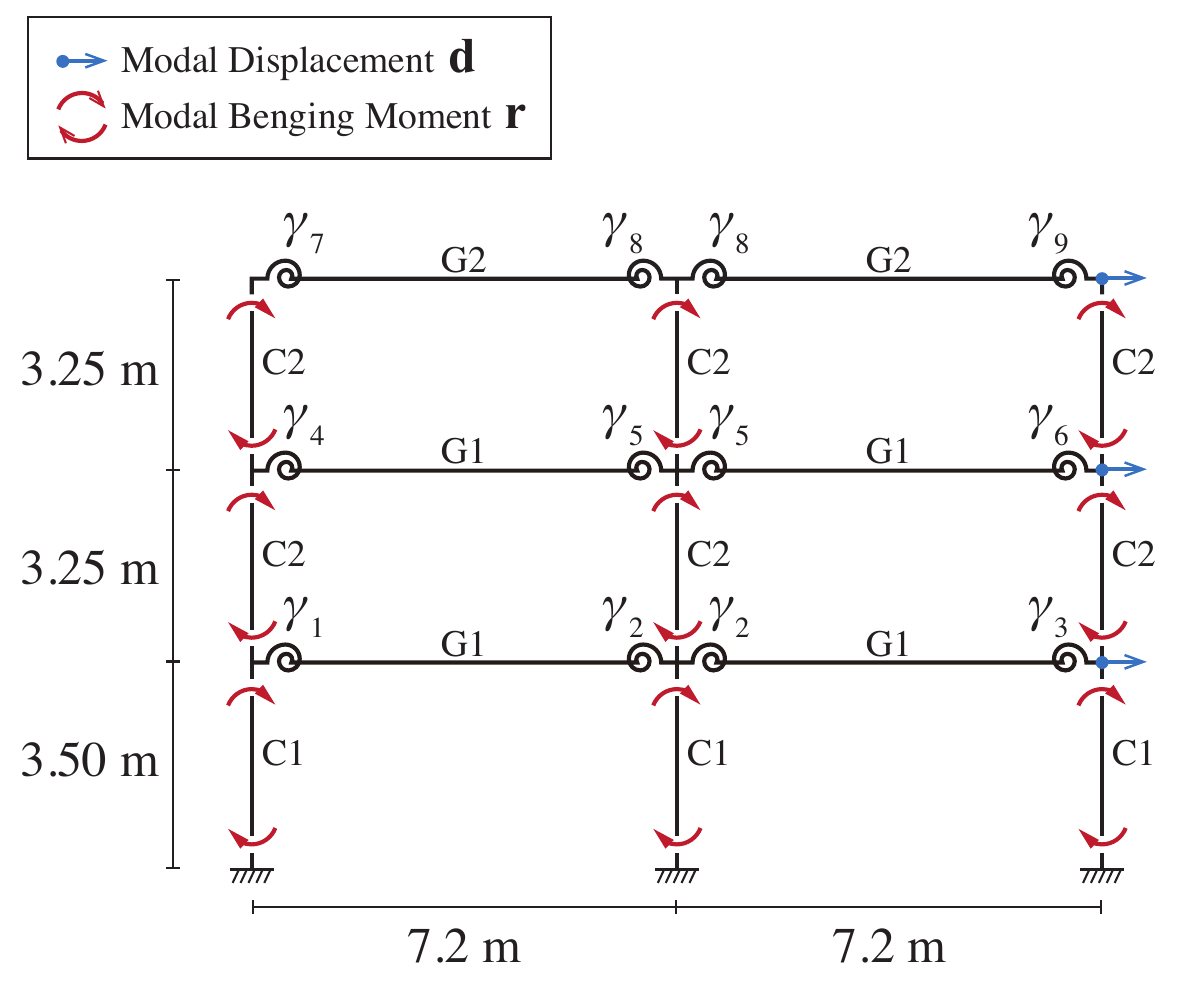}
    \caption{Three-story two-bay planar frame targeted in the numerical example.}
    \label{fig:2bay3story}
\end{figure}

\begin{table}[!t]
    \centering
    \caption{Assumed members of the two-bay three-story planar frame.}
    \label{tab:members}
    \fontsize{8truept}{10truept}\selectfont
    \begin{tabular*}{\textwidth}{@{\extracolsep{\fill}}llrr}
        \toprule
        Member & Cross section & Cross-sectional area $A$ (mm$^2$) & Second moment of the area $I$ (mm$^4$) \\
        \midrule
        G1 & H--500$\times$200$\times$10$\times$16 & $1.123 \times 10^2$ & $4.680 \times 10^8$ \\
        G2 & H--400$\times$200$\times$ 8$\times$13 & $8.337 \times 10^1$ & $2.350 \times 10^8$ \\
        C1 & BOX--300$\times$300$\times$19 & $2.043 \times 10^2$ & $2.620 \times 10^8$ \\
        C2 & BOX--300$\times$300$\times$12 & $1.345 \times 10^2$ & $1.830 \times 10^8$ \\
        \bottomrule
    \end{tabular*}
\end{table}

Consider a three-story, two-bay steel moment resisting frame as shown in Figure~\ref{fig:2bay3story}.
The Young's modulus and specific gravity are determined as $E = 2.05 \times 10^5$ (N/mm$^2$) and $\rho = 7.85$, respectively.
The shear deformation is ignored (i.e., the shear modulus $G = \infty$).
The cross-sectional properties of the beams and columns are listed in Table~\ref{tab:members}.
We assume a uniformly distributed load of 43.75 kN/m acting on the beams other than self-weight, and assign equivalent lumped masses to each node.

Each beam is equipped with semi-rigid rotational springs at both ends.
To avoid severe ill-posedness, the two springs connected to each intermediate column are constrained to share the same rotational stiffness.
Thus, we have a total of nine stiffness parameters, denoted by $k_i, ~ i = 1, ..., 9$.
These are transformed into a \textit{fixity factor} \cite{dhillonBS1990},
\begin{align}
    \gamma_i = \left( 1 + \frac{3 EI/L}{k_i} \right)^{-1},
\end{align}
where $EI$ and $L$ are the bending stiffness and length of each beam.
The locations of the beam ends associated with $\gamma_1,...,\gamma_9$ are indicated in Figure~\ref{fig:2bay3story}.
The factor satisfies $\gamma_i \in [0, 1]$; it represents a perfectly rigid joint when $\gamma_i = 1$ and a pinned joint when $\gamma_i = 0$.
The system parameters to be updated are therefore $\bm{\theta} := \{\gamma_i\}_{i = 1}^9 \in \mathbb{R}^9$;
the other properties are assumed to be known.

\begin{figure}[!t]
    \centering
    \includegraphics[width=1.0\linewidth]{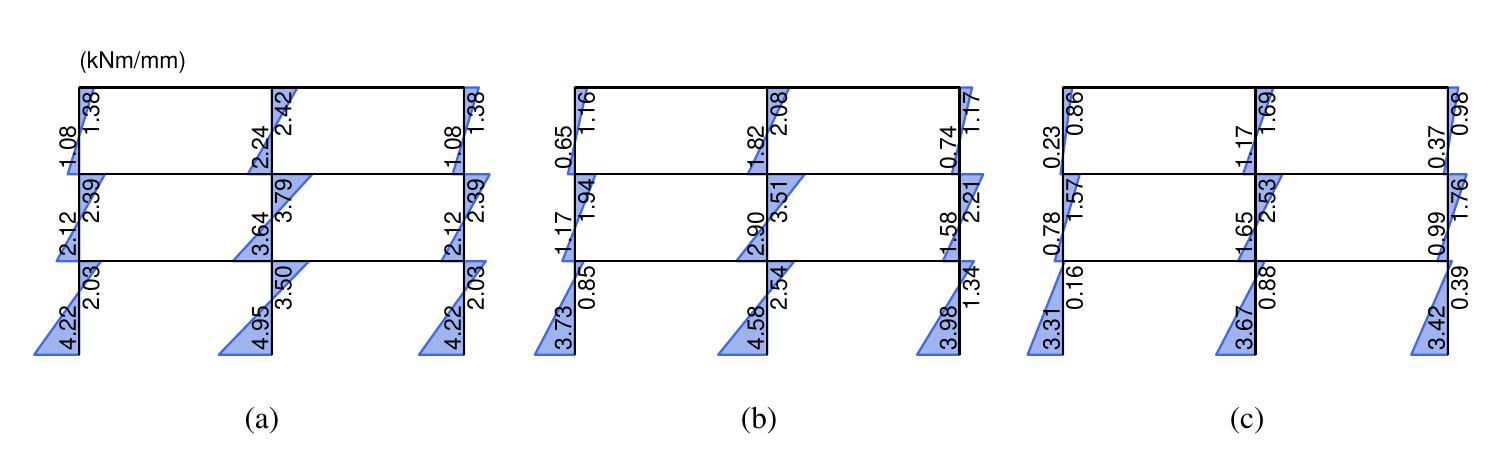}
    \caption{Normalized modal bending moment diagrams for three damage states: (a) intact ($k=1$); (b) moderate ($k=2$); (c) severe ($k=3$).}
    \label{fig:num-mbm}
\end{figure}

As shown in Figure~\ref{fig:2bay3story}, MBM is assumed to be measured at 18 locations (the top and bottom of each column) and MD is assumed to be measured at three locations (one for each floor), denoted by $\mathbf{r} \in \mathbb{R}^{18}$ and $\mathbf{d} \in \mathbb{R}^3$, respectively.
We then assume the nMBM observations, $\mathcal{X} := \{\overline{\mathbf{r}}_n : \overline{\mathbf{r}}_n\ = \mathbf{r}_n / \|\mathbf{d}_n\| \in \mathbb{R}^{18}\}$.
To synthesize the dataset $\mathcal{X}$, we adopt the following generative model,
\begin{alignat}{2}
     \overline{\mathbf{r}}_n ~ &\sim ~ \mathcal{N}(\cdot \mid h(\bm{\theta}_n), \sigma^2   \mathbf{I}), \quad & n = 1, ..., N, \\
    \bm{\theta}_n ~ &\sim ~ \mathcal{N}(\cdot \mid \bm{\mu}_{c_n}, \sigma^2_0 \mathbf{I}), \quad & n = 1, ..., N,
\end{alignat}
where $h$ denotes a function that takes $\bm{\theta}$ as inputs and outputs an nMBM vector via an eigenvalue analysis.
$\sigma$ denotes the standard deviation of the observation errors. Three noise levels are considered: $\sigma \in \{0.05, 0.10, 0.20\}$ (kNm/mm).
$\sigma_0$ denotes the standard deviation that controls the stochasticity of $\bm{\theta}$ and is fixed at $\sigma_0 = 0.02$.
Three classes are assumed, $c_n \in \{1, 2, 3\}$, with five observations deterministically assigned to each class; that is,
\begin{align}
    c_n =
    \begin{cases}
        1 & \quad \text{if} \quad  1 \leq n \leq 5  \\
        2 & \quad \text{if} \quad  6 \leq n \leq 10 \\
        3 & \quad \text{if} \quad 11 \leq n \leq 15,
    \end{cases}
\end{align}
resulting in a total of $N = 15$ observations.
Table~\ref{tab:centroids} lists the ground truth values of $\bm{\mu}_k = \{\mu_{ki}\}_{i=1}^9$, where $\mu_{ki}$ denotes the mean value associated with $\gamma_i$ for class $k$.
Class $k = 1$ represents the intact state, in which all nodes are assumed to have rigid beam ends with a fixity factor of $\gamma_i = 0.9$.
Classes $k = 2$ and $k = 3$ correspond to moderate and severe damage states, respectively.
In both cases, moderate-to-severe damages ($\gamma_i < 0.5$) are assumed to be concentrated in the first story (the second floor).
Figure~\ref{fig:num-mbm} presents the nMBM diagrams for the three damage cases, computed under ideal conditions (no parameter uncertainty and no measurement noise).
Comparing the severe damage state (c) to the intact case (a) shows markedly reduced nMBM values in columns adjacent to severely damaged beam ends (i.e., the first-story columns), underscoring the strong sensitivity of nMBM to localized damage.

\begin{table}[!t]
    \centering
    \caption{Ground truth values of mean vectors, $\bm{\mu}_k$.}\label{tab:centroids}
    \fontsize{8truept}{10truept}\selectfont
    \begin{tabular*}{\textwidth}{@{\extracolsep{\fill}}llrrrrrrrrr}
        \toprule
        Damage states & Class, $k$ & $\mu_{k1}$ & $\mu_{k2}$ & $\mu_{k3}$ & $\mu_{k4}$ & $\mu_{k5}$ & $\mu_{k6}$ & $\mu_{k7}$ & $\mu_{k8}$ & $\mu_{k9}$ \\ \midrule
        Intact   & 1 & 0.9 & 0.9 & 0.9 & 0.9 & 0.9 & 0.9 & 0.9 & 0.9 & 0.9 \\
        Moderate & 2 & 0.4 & 0.7 & 0.6 & 0.7 & 0.9 & 0.8 & 0.9 & 0.9 & 0.9 \\
        Severe   & 3 & 0.2 & 0.3 & 0.3 & 0.5 & 0.6 & 0.6 & 0.7 & 0.8 & 0.8 \\
        \bottomrule
    \end{tabular*}
\end{table}

\subsection{Bayesian inference}

We then apply DP-HBMU to the synthetic data $\mathcal{X} = \{\overline{\mathbf{r}}_n\}_{n=1}^{15}$.
For tractability in terms of Gaussian mixtures, $\bm{\theta}$ with a bounded support $[0, 1]$ is transformed into $\tilde{\bm{\theta}} = \Phi^{-1}(\bm{\theta})$, where $\Phi$ denotes the cumulative distribution function (CDF) of the standard normal distribution.
The hyperparameters of the Gamma distributions $\mathrm{Ga}(\cdot \mid a, b)$ are specified such that the prior mean $a/b$ represents the anticipated value of the parameter, while the coefficient of variation (CoV) $1/\sqrt{a}$ reflects the degree of confidence in this prior belief.
For the cluster precision parameter, $\tau$, we set $(a_\tau, b_\tau) = (2, 0.02)$.
This choice yields a prior mean of 100, which corresponds to a within-cluster standard deviation of approximately $100^{-1/2}=0.1$, and a CoV of $1/\sqrt{2} \simeq 0.7$.
For the precision parameter associated with the observation errors, $\beta$, we set $(a_\beta, b_\beta) = (2, 0.02)$.
This choice yields a prior mean of 100, which corresponds to the assumption that observations $\overline{\mathbf{r}}_n$ are subject to measurement errors with a standard deviation of 0.1 (kNm/mm).
For the CRP scale parameter $\alpha$, we set $(a_\alpha, b_\alpha) = (1, 1)$, yielding a prior mean of 1 and CoV of 1.
This choice corresponds to a relatively wide prior, avoiding undue bias in the number of clusters.
The hyperparameters of the Gaussian prior $\mathcal{N}(\bm{\mu}_k \mid \bm{\mu}_0, (\rho\tau)^{-1}\mathbf{I})$ are specified as $\bm{\mu}_0 = \mathbf{0}$ and $\rho = 0.05$.
The parameter $\rho$ is a key factor governing the degree of separation between clusters.
Although smaller values of $\rho$ improve cluster separability, they limit knowledge transfer across clusters and may lead to less stable inference in practice.
$\rho$ should be chosen in a problem-dependent manner, considering the expected number of observations per cluster.

We additionally examine a non-hierarchical case (termed non-hierarchical BMU) to confirm the effect of introducing the hierarchical architecture.
The generative model of each observation $\overline{\mathbf{r}}_n$ is given by
\begin{equation}
    \label{eq:gen-nhbmu}
    \begin{aligned}
        \overline{\mathbf{r}}_n ~ &\sim ~ \mathcal{N}(\cdot \mid h(\bm{\theta}_n), \beta^{-1} \mathbf{I}), \\
        \bm{\theta}_n ~ &\sim ~ \mathcal{U}(\cdot \mid [0, 1]^D), \\
        \beta_n ~ & \sim ~ \mathrm{Ga}(\cdot \mid a_\beta, b_\beta),
    \end{aligned}
\end{equation}
where $\mathcal{U}(\cdot \mid \mathcal{S})$ denotes a uniform distribution over a set $\mathcal{S}$.
In this case, we estimate $(\bm{\theta}_n, \beta_n)$ for each $\overline{\mathbf{r}}_n$ via the Metropolis-within-Gibbs sampler that employs a pCN proposal for $\tilde{\bm{\theta}}_n := \Phi^{-1}(\bm{\theta}_n)$.
The step size $s$ is adaptively tuned during the burn-in period according to the updating rule in Eq.~(\ref{eq:update_s}), with the target acceptance rate set to $\lambda^\ast = 0.4$, reflecting the absence of a hierarchical structure.
The hyperparameters of the Gamma prior are set as $(a_\beta, b_\beta) = (2, 0.02)$, consistent with those in the hierarchical case.

As a performance metric to compare DP-HBMU with non-hierarchical BMU, we use the log posterior probability at the true parameters.
Let $\bm{\theta}_n^\ast$ denote the true parameters used to synthesize the observation $\mathbf{x}_n$.
As the joint posterior $p(\{\bm{\theta}_n\} \mid \mathcal{X})$ is difficult to evaluate due to its high dimensionality, we approximate it by the sum of the log marginal posteriors (hereafter termed SLMP),
\begin{align}
    \text{SLMP} := \sum_{n=1}^{N} \sum_{i=1}^{D} ~ \log p(\theta_{ni} = \theta_{ni}^\ast \mid \mathcal{X}),
\end{align}
where $p(\theta_{ni} \mid \mathcal{X})$ can be computed using kernel density estimations.

\begin{table}[!t]
    \centering
    \caption{Performance metrics, SLMP values, for both hierarchical and non-hierarchical methods: the summary of ten independent runs.}\label{tab:num-results}
    \fontsize{8truept}{10truept}\selectfont
    \begin{tabular*}{\textwidth}{@{\extracolsep{\fill}}rrrrr}
        \toprule
        $\sigma$ (kNm/mm) & \multicolumn{2}{c}{DP-HBMU (Proposed)} & \multicolumn{2}{c}{Non-hierarchical BMU} \\
        \cmidrule{2-3} \cmidrule{4-5}
        ~ & mean & std. & mean & std. \\
        \midrule
        $0.05$ & $9.66$ & $0.97$ &  $6.56$ & $0.29$ \\
        $0.10$ & $4.79$ & $1.32$ &  $2.47$ & $0.57$ \\
        $0.20$ & $0.65$ & $1.14$ & $-3.19$ & $2.15$ \\
        \bottomrule
    \end{tabular*}
\end{table}

\begin{figure}[!t]
    \centering
    \includegraphics[width=0.8\textwidth]{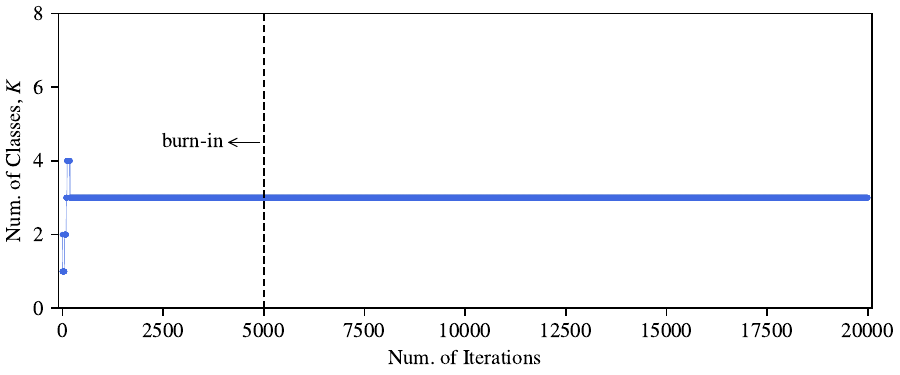}
    \caption{Estimated number of classes $K$ during the MCMC iterations.}
    \label{fig:num-classes}
\end{figure}

\begin{figure}[!t]
    \vspace{4truemm}
    \centering
    \includegraphics[width=1.0\textwidth]{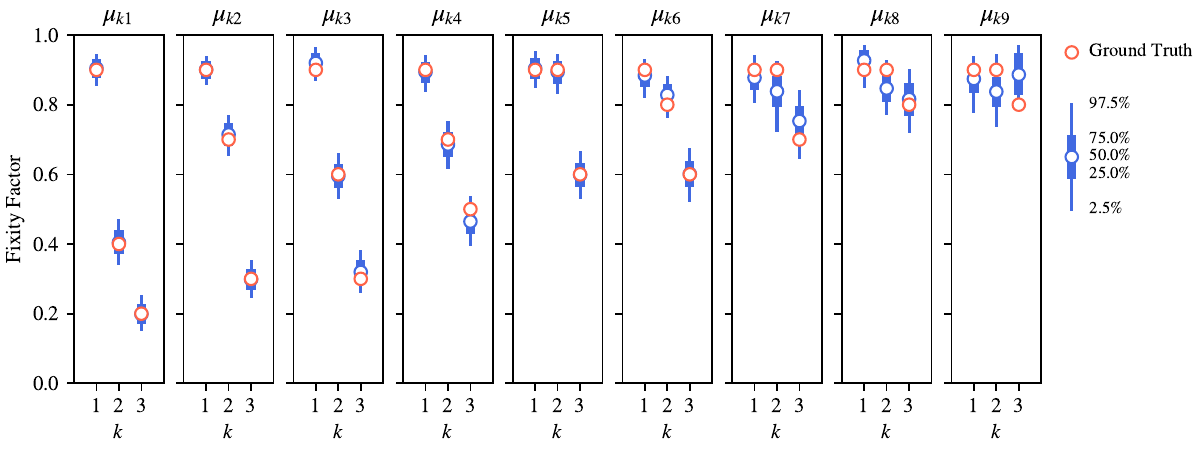}
    \caption{Posterior samples of $\{\bm{\mu}_k\}$ conditional on $K = 3$.}
    \label{fig:num-mus}
\end{figure}

For both cases, we perform the Metropolis-within-Gibbs sampling with 20000 iterations, discarding the first $T_\mathrm{burn} = 5000$ iterations as a burn-in period.
The convergence is assessed by visually examining the trace plots of the parameters and monitoring the stability of the class assignments $\{c^{(t)}_n\}$.
Table~\ref{tab:num-results} shows the SLMP scores of DP-HBMU and non-hierarchical BMU for different levels of noise $\sigma$.
For each $\sigma$, the table presents the mean and standard deviation of the ten independent runs.
The proposed method consistently outperforms the non-hierarchical BMU for all cases.

\begin{figure}[!t]
    \centering
    \includegraphics[width=1.0\textwidth]{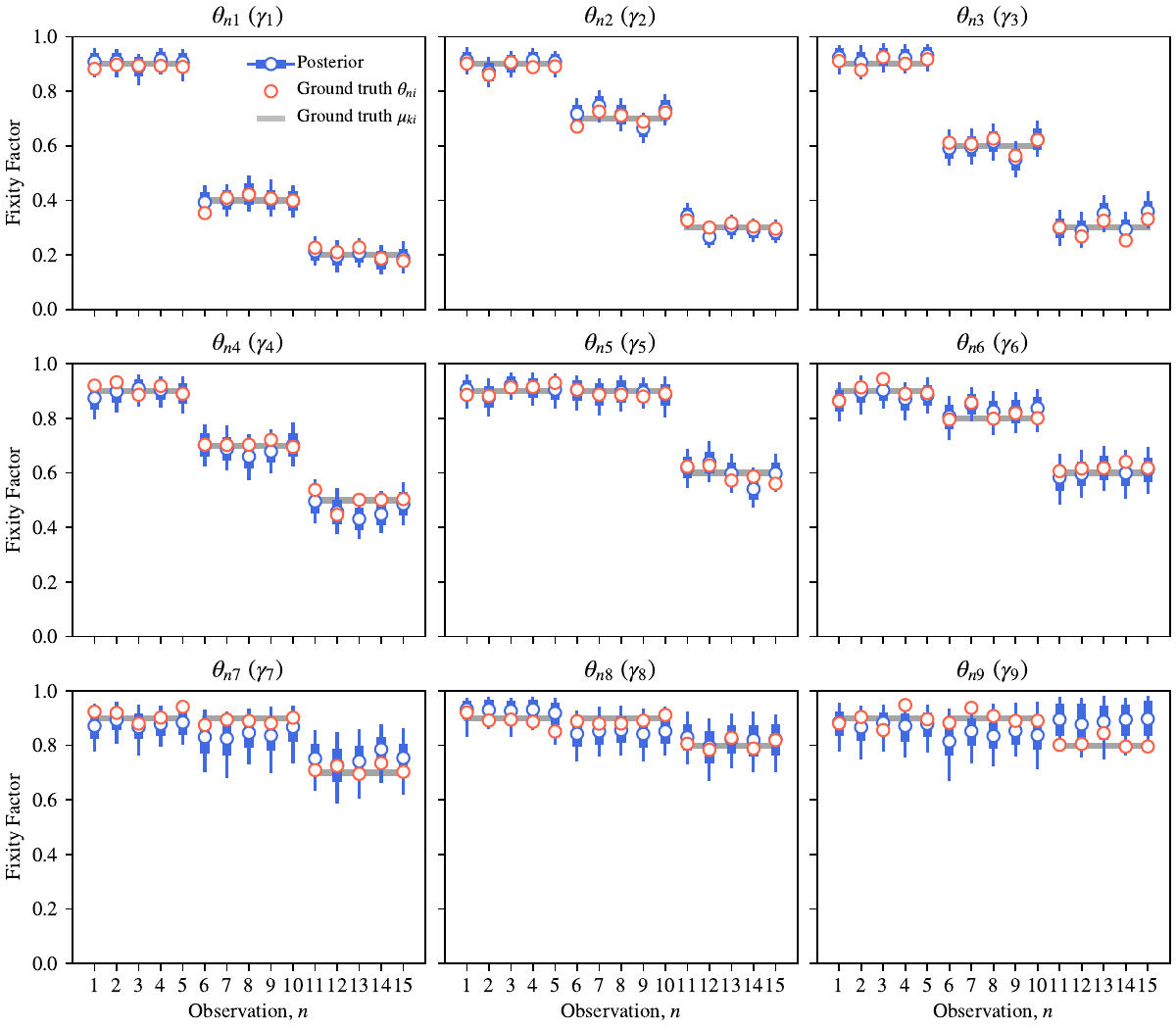}
    \caption{Posterior samples of $\{\bm{\theta}_n\}$.}
    \label{fig:num-xxs}
\end{figure}

To further illustrate the effectiveness of the proposed algorithm, we hereafter focus on a simulation with $\sigma = 0.1$.
Figure~\ref{fig:num-classes} shows the trace of $K$ throughout the MCMC iterations, which converges to the expected number, $K = 3$.
Figure~\ref{fig:num-mus} summarizes the posterior samples of the mean vector $\bm{\mu}_k = \{\mu_{ki}\}_{i=1}^9$ conditional on $K=3$.
For almost all components $\mu_{ki}$, the posterior medians closely align with the ground truth values, and the associated uncertainty intervals are generally tight.
We note that minor discrepancies exist between the posterior medians of $\mu_{29}$ and $\mu_{39}$ and their ground-truth values.
These likely stem from the isotropic Gaussian noise assumption on $\overline{\mathbf{r}}_n$, which reduces the signal-to-noise ratio in the upper stories where the response amplitudes are smaller.
Figure~\ref{fig:num-xxs} showcases the posterior samples of $\bm{\theta}_n, ~ n = 1, ..., 15$.
For most components $\theta_{ni}$, the posterior medians closely align with the ground truth values.
Notably, the inferred posteriors track even subtle variations within a cluster, especially evident for $\theta_{n1}, ~ n = 11, ..., 15$.

In summary, the inference results clearly show that the proposed method, DP-HBMU, classifies damage states included in the datasets without pre-specifying the number of clusters.
Moreover, owing to its hierarchical structure, it yields more robust parameter estimation with narrower uncertainty intervals.

\section{Experimental Validation}\label{sec:exp}

\subsection{Target structure and data}\label{sec:exp-target}

\begin{figure}[!t]
    \centering
    \includegraphics[width=0.40\textwidth]{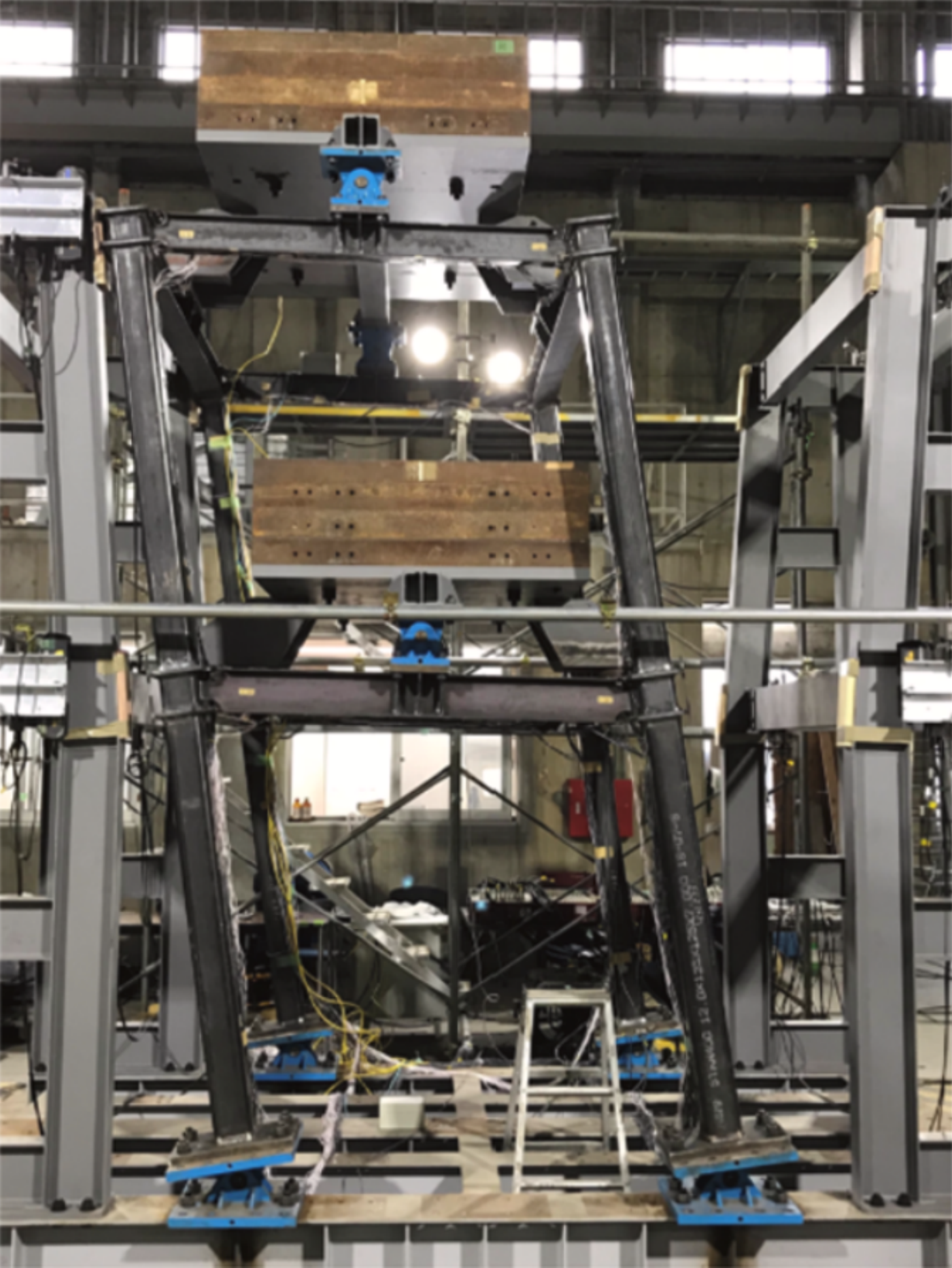}
    \caption{Target specimen in experimental validation: south-side view after completion of loading tests (from Iyama et al.~\cite{iyamaJ2021ES}).}
    \label{fig:exp-specimen}
\end{figure}

\begin{figure}[!t]
    \vspace{6truemm}
    \centering
    \includegraphics[width=1.00\linewidth]{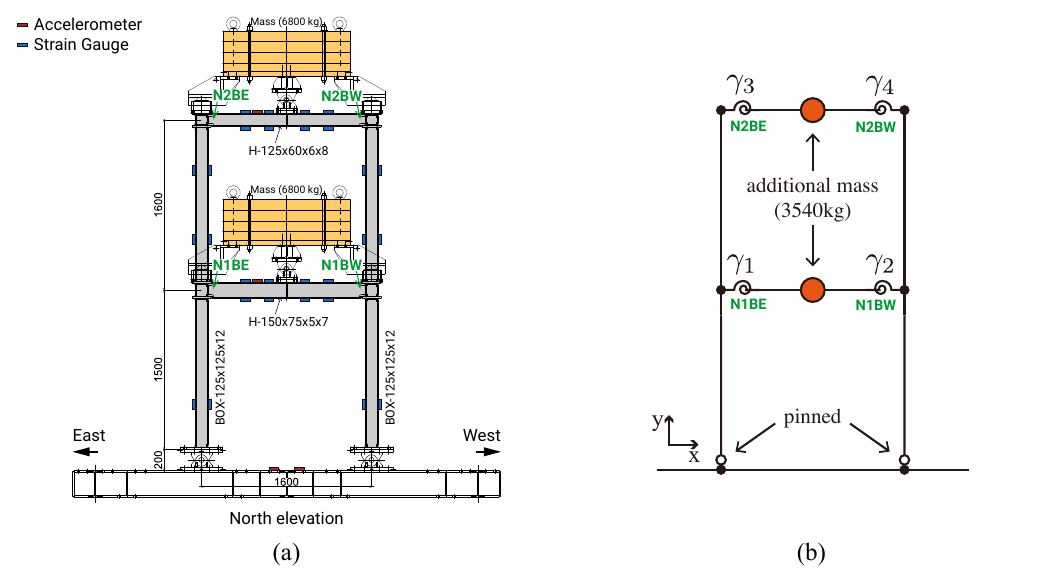}
    \caption{Target structure in experimental validation \cite{iyamaJ2021ES}:
    (a)~north elevation; (b)~finite element model (planar moment frame with semi-rigid beam ends and pinned column bases).}
    \label{fig:exp-frame}
\end{figure}

In this section, we apply DP-HBMU to experimental data \cite{iyamaJ2021ES} within a framework of stress resultant-based model updating.
The target structure is a two-story, one-bay by one-bay steel moment frame with pin joints at the column bases, as shown in Figure~\ref{fig:exp-specimen}.
Each floor carried an additional mass of 6800 kg, attached to the beams by pin joints.
The specimen experienced dynamic loadings only in the east-west (EW) direction.
We hereafter focus on the north frame, while considering both the north and south frames for modal identification (i.e., computation of nMBM values).
Figure~\ref{fig:exp-frame} illustrates the north elevation of the frame with sensor configuration considered in this study.
Accelerometers were installed on each beam that spans the EW axis as well as on the base.
Strain gauges were attached to the top and bottom edges of the selected beam and column sections.
In the north frame, we focus on the rotational stiffness of the four beam-ends, termed N1BE, N1BW, N2BE, and N2BW, as indicated in Figure~\ref{fig:exp-frame} (b).

Dynamic strain measurements from these gauges are converted to bending moment responses via the following relationship,
\begin{align}
    M(t) = \frac{\varepsilon_\mathrm{t}(t) - \varepsilon_\mathrm{b}(t)}{2} EZ,
\end{align}
where $M(t)$ denotes the bending moment at time $t$, and $\varepsilon_\mathrm{t}(t)$ and $\varepsilon_\mathrm{b}(t)$ denote the strains measured at the top and bottom edges of the cross section considered at time $t$.
$E$ is the Young's modulus and $Z$ is the elastic section modulus; both are taken at their nominal values.
Assuming a linear bending‐moment distribution along each member, we interpolate the bending moments between instrumented sections to estimate the ones at the beam and column ends.
The bending moments at the pinned column bases are assumed to be zero.

In the suite of loading tests, recorded seismic ground motion (JMA Kobe NS) and broadband random noise were alternately applied to the specimen.
The amplitude of the JMA Kobe NS wave was incrementally scaled to reproduce progressive damage states.
The present study focuses on seven random-wave tests, R3, R4, ..., R9, as listed in Table~\ref{tab:exp-tests}.
The table also shows the damage state that was observed at the four beam ends prior to each loading.
Fracture was first observed at N1BE in R7 and then at N1BW in R8, suggesting that the test series can be classified into three damage states: (i) no significant damage, (ii) severe damage at a single beam end, and (iii) severe damage at two beam ends.
To increase the number of observations, each 60 s test record is partitioned into two 30 s segments, which are treated as individual datasets.
This yields 14 datasets in total, denoted by R3a, R3b, ..., R9a, R9b.

Figure~\ref{fig:exp-specimen} (b) illustrates the assumed FE model composed of beam elements.
Pin supports are assumed at the column bases, and rotational springs are placed at beam ends to represent the semi-rigid connections.
The rotational stiffnesses of these springs are transformed into fixity factors, $\gamma_1, ..., \gamma_4$, which serve as the system parameters to be updated, $\bm{\theta} \in \mathbb{R}^4$.

\begin{table}[!t]
    \centering
    \fontsize{9truept}{11truept}\selectfont
    \caption{Observed damage states at beam ends of the north frame in dynamic loading tests \cite{iyamaJ2021ES}.} \label{tab:exp-tests}
    \begin{tabular*}{\textwidth}{@{\extracolsep{\fill}}lllll}
        \toprule
        ~ & \multicolumn{4}{c}{Damage states} \\
        \cmidrule{2-5}
        Test & N1BE & N1BW & N2BE & N2BW \\ \midrule
        R3 & --                 & Wrinkle           & --        & --    \\ 
        R4 & Wrinkle            & Crack             & --        & --    \\ 
        R5 & Crack              & Crack             & --        & --    \\
        R6 & Extending          & Extending         & --        & --    \\ 
        R7 & \textbf{Fracture}  & Penetration       & Wrinkle   & --    \\ 
        R8 & \textbf{Fracture}  & \textbf{Fracture} & Crack     & --    \\ 
        R9 & \textbf{Fracture}  & \textbf{Fracture} & Extending & Crack \\ 
        \bottomrule
    \end{tabular*}
\end{table}

\subsection{Modal identification}

\begin{figure}[!t]
    \centering
    \includegraphics[width=1.0\textwidth]{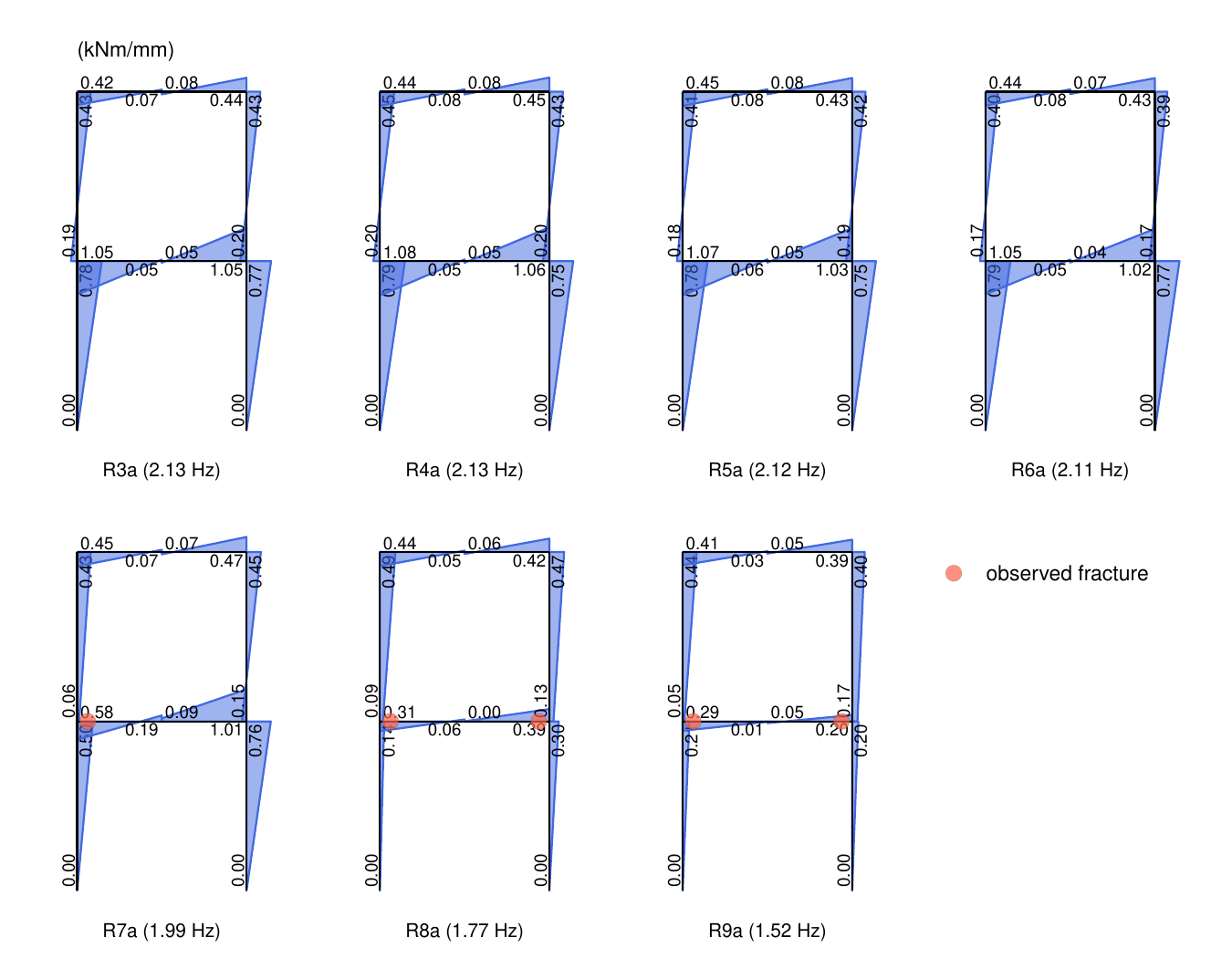}
    \caption{Normalized modal bending moment distributions evaluated via system identification.}
    \label{fig:mbm_1}
\end{figure}

The nMBM can be computed from the measured strain and acceleration as follows.
We first perform multi-input multi-output (MIMO) system identification, using the base accelerations as the inputs and acceleration response at each floor and bending moment responses at beam and column ends as the outputs.
The MOESP method \cite{verhaegenM1992a}, a class of subspace state space system identification (4SID), is employed with a system order of four, which suggests two pairs of complex-conjugate modes.
The number of block rows in a block Hankel matrix is set to 50.
This results in modal frequency $\omega_i$, mode shapes in terms of acceleration (modal acceleration) $\mathbf{a}_i$, and an MBM vector $\mathbf{r}_i$ for each mode $i$.
The MD and nMBM are then obtained as $\mathbf{d}_i = - \mathbf{a}_i/\omega_i^2$ and $\overline{\mathbf{r}}_i = \mathbf{r}_i / \|\mathbf{d}_i\|$ \cite{yaoyamaT2024a,yaoyamaT2024b,yaoyamaT2024isrerm}.
In the Bayesian inference, we use the nMBM vector for the first mode only,
as that for the second mode was observed to be less stable, likely due to its higher sensitivity to measurement noise.
Figure~\ref{fig:mbm_1} showcases the first-mode nMBM distribution diagrams with the corresponding modal frequencies for selected tests.
From R3a to R6a, the natural frequency remains nearly constant at about 2.1 Hz, and the nMBM presents no significant change.
From R6a to R7a, the natural frequency decreases by about 5\% and the nMBM at N1BE drops from 0.58 to 0.31 (kNm/mm).
From R7a to R8a, the natural frequency further decreases by about 11\% and nMBM at N1BW drops from 1.01 to 0.39 (kNm/mm).
These reflect the sequentially observed fractures at N1BE and N1BW, suggesting that the nMBM is a sensitive indicator of localized stiffness damages.

\begin{figure}[!t]
    \vspace{6truemm}
    \centering
    \includegraphics[width=0.8\textwidth]{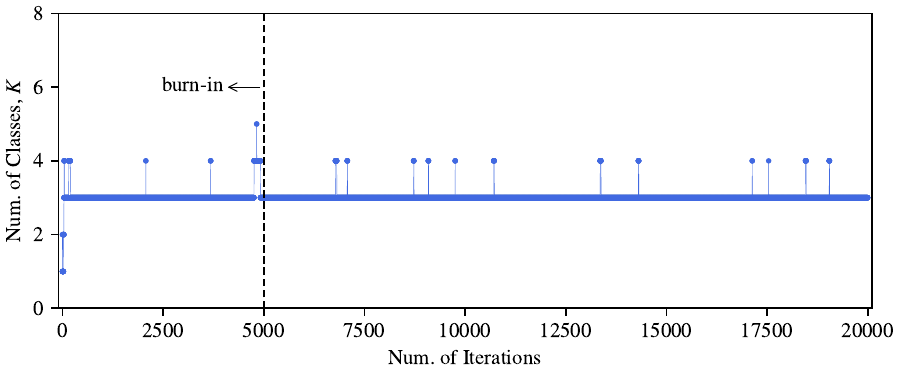}
    \caption{Estimated number of classes $K$ during the MCMC iterations.}\label{fig:exp-classes}
\end{figure}

\begin{figure}[!t]
    \centering
    \includegraphics[width=1.0\textwidth]{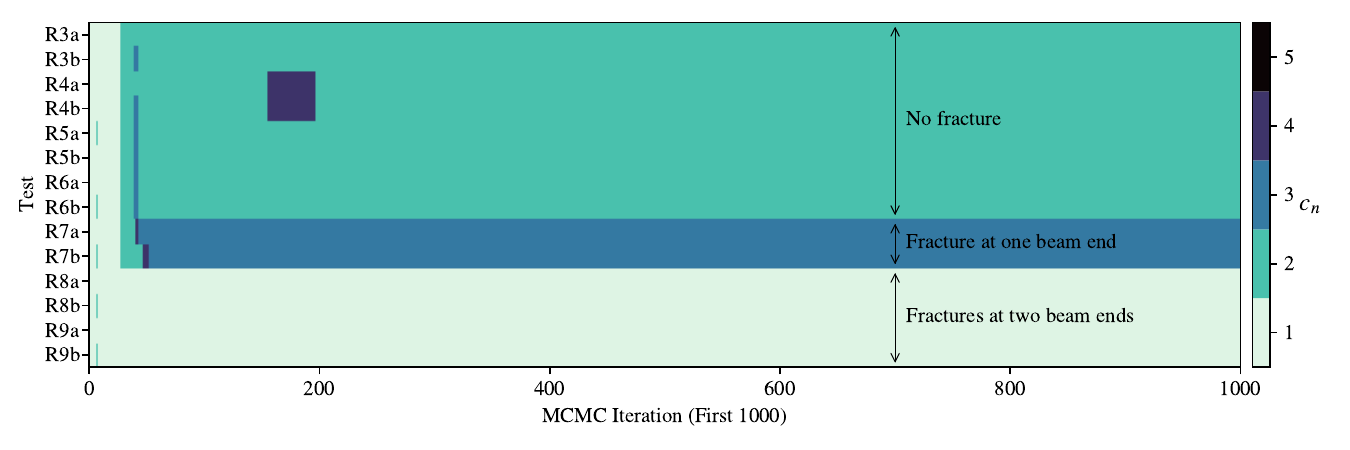}
    \caption{Cluster assignments over MCMC iterations (only the first 1000 iterations shown).}\label{fig:exp-clusters}
\end{figure}

\begin{figure}[!t]
    \vspace{6truemm}
    \centering
    \includegraphics[width=0.7\textwidth]{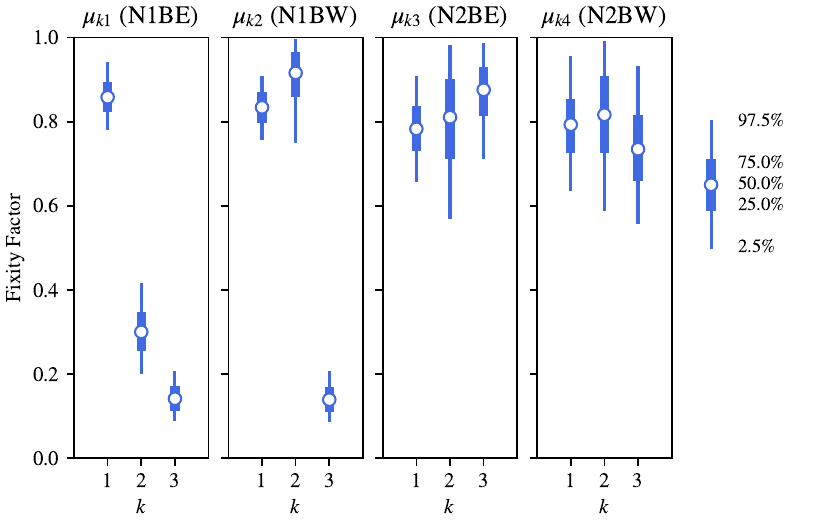}
    \caption{Posterior samples of $\{\bm{\mu}_k\}$.}\label{fig:exp-mus}
\end{figure}

\begin{figure}[!t]
    \centering
    \includegraphics[width=1.0\textwidth]{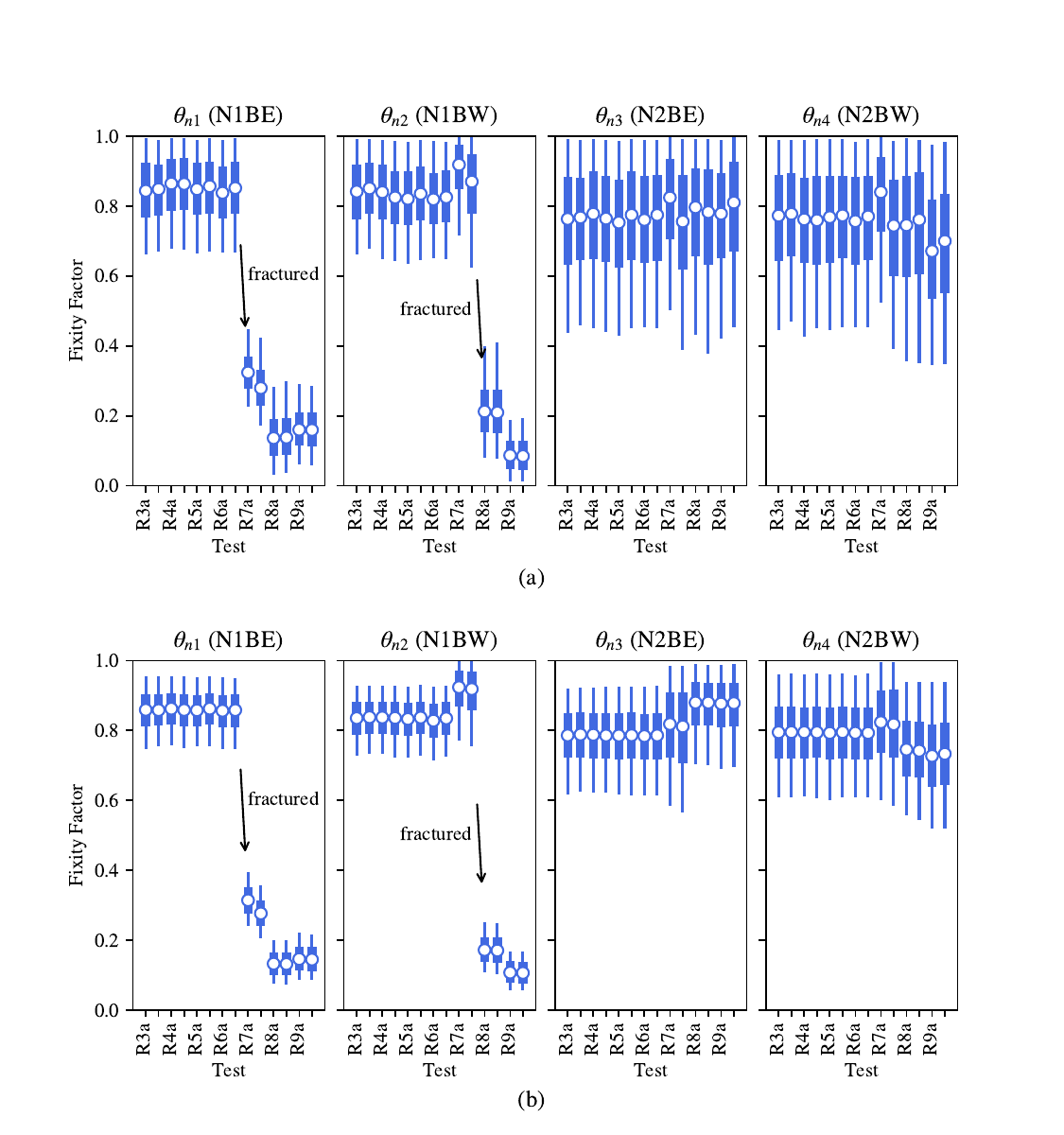}
    \caption{Posterior samples of fixity factors $\gamma_i, ~ i = 1, ..., 4$ for all tests: (a) non-hierarchical BMU; (b) DP-HBMU (Proposed).}\label{fig:exp-xxs}
\end{figure}

\subsection{Model updating}

A set of the identified nMBM vectors is considered as ``observations'', i.e., $\mathcal{X} := \{\overline{\mathbf{r}}_n\}_{n=1}^{14}$, for model updating.
The hyperparameters of the Metropolis-within-Gibbs sampler are set in the same manner as in the numerical example: that is, $\rho = 0.05$, $\bm{\mu}_0 = \mathbf{0}$, $(a_\beta, b_\beta) = (2, 0.02)$, $(a_\tau, b_\tau) = (2, 0.02)$, $(a_\alpha, b_\alpha) = (1, 1)$.
The target acceptance rate $\lambda^\ast$ for the pCN algorithm is set to $0.8$.
We also consider a non-hierarchical baseline case defined in Eq.~(\ref{eq:gen-nhbmu}) with $(a_\beta, b_\beta) = (2, 0.02)$ and $\lambda^\ast = 0.4$.
For both cases, we perform the Metropolis-within-Gibbs sampler for 20000 iterations and discard the first $T_\mathrm{burn} = 5000$ iterations as burn-in.

Figure~\ref{fig:exp-classes} shows the trace of the number of classes $K$ across the MCMC iterations.
This converges to $K = 3$, the expected number of damage states.
Figure~\ref{fig:exp-clusters} presents the trace of the cluster assignments $\{c_n\}$ specifically for the first 1000 iterations.
The assignments gradually converge into the expected three classes---no fracture (R3a--R6b), one beam-end fracture (R7a, R7b), and two beam-end fractures (R8a--R9b).

Figure~\ref{fig:exp-mus} presents posterior samples of the class mean vectors $\{\bm{\mu}_k\}$ for each beam end conditional on $K = 3$.
Since label switching occurs during MCMC sampling, the class labels are relabeled using the procedure described in Section~\ref{sec:relabeling}.
For visualization purposes, the relabeled classes are further ordered to reflect the progression of damage.
In class $k = 1$, the medians across all beam ends lie between 0.7 and 0.9.
In class $k = 2$, the median for N1BE decreases to roughly 60\% of its initial value, whereas those for the other beam ends deviate by no more than about 20\%, reflecting the observed fracture at N1BE.
In class $k = 3$, the median for N1BW drops by more than 80\%, whereas those for N2BE and N2BW remain above 0.6, corresponding to the observed fracture at N1BW.

Figure~\ref{fig:exp-xxs} presents posterior samples of the fixity factors $\{\bm{\theta}_n : \bm{\theta}_n = \{\theta_{n1}, ..., \theta_{n4}\}, n = 1, ..., 14\}$, in comparison with the case of non-hierarchical BMU.
In both sub-figures (a) and (b), the stiffness decreases due to the fractures are successfully reflected.
In both cases, the inferred posteriors correctly capture the stiffness decreases at N1BE and N1BW associated with fracture.
A comparison of the two cases reveals that, in (b) (DP‑HBMU), the median estimates evolve more smoothly and the associated uncertainty ranges are substantially narrower than in (a) (non‑hierarchical BMU), particularly at N2BE and N2BW. These findings demonstrate that the DP‑HBMU identifies the distinct damage states present in the dataset, and achieves a pronounced reduction in parameter uncertainty by leveraging its hierarchical structure to transfer knowledge within each state.

\section{Conclusions}\label{sec:conclusions}

This paper has proposed a hierarchical Bayesian model updating framework that employs a Dirichlet process mixture prior on model parameters (DP-HBMU).
The DP-HBMU framework can cluster observations in the structural parameter space according to damage states without pre-specifying the number of clusters.
For the inference algorithm, we devised a Metropolis-within-Gibbs sampler that embeds Metropolis–Hastings updates to sample the structural parameters for each observation, addressing intractable conditionals arising from the FE simulator.
Numerical and experimental examples demonstrated the applicability of DP-HBMU to damage localization problems, especially focusing on a stress resultant-based model updating of frame structures. 
In the numerical example, we examined a two-bay, three-story planar frame with semi-rigid beam ends under three damage states including intact, moderate, and severe states.
DP-HBMU successfully recovered these three states.
The inferred posterior of the rotational stiffness at beam ends aligned closely with the ground-truth values and exhibited narrower uncertainty than a non-hierarchical baseline case, suggesting the effectiveness of the proposed algorithm.
In the experimental validation, we applied DP-HBMU to dynamic loading test data on a two-story, one-bay by one-bay steel moment frame.
The proposed method classified 14 datasets into three clusters corresponding to the damage states assumed from the observed beam-end fractures.
The inferred posterior revealed more stable variation within each damage state and markedly reduced uncertainties compared to a non-hierarchical baseline case.
These examinations indicate that the proposed DP-HBMU framework combines probabilistic FE model updating with damage classification without prior knowledge of the number of damage states included in the dataset.

As future work, we will relax the shared-covariance assumption by adopting component-specific covariance structures, such as $\tau_k^{-1}\mathbf{I}$ or full covariance matrices $\mathbf{\Sigma}_k$, and to evaluate the proposed framework under more realistic scenarios, including heterogeneous cluster covariances and unequal cluster sizes.
Furthermore, we plan to devise scalable inference schemes that remain effective on larger datasets (e.g., $N > 50$) by introducing the surrogate models for the simulator or likelihood functions.
We will also examine the applicability of the proposed framework to high dimensional model parameters $\bm{\theta}$, particularly with respect to posterior complexity and multimodality, which can affect the efficiency of the overall Metropolis-within-Gibbs sampler.

\section*{Declaration of generative AI and AI-assisted technologies in the writing process}

During the preparation of this work the authors used ChatGPT in order to improve the readability and language of the manuscript.
After using this tool, the authors reviewed and edited the content as needed and take full responsibility for the content of the published article.

\appendix

\section{Split--merge algorithm to update cluster assignments}\label{sec:app}

Following Jain and Neal~\cite{jainS2004}, we update the class labels $\{c_n\}$ by a restricted split--merge Metropolis--Hastings move.
Let $i$ and $j$ denote two distinct indices drawn uniformly at random from $\{1,\ldots,N\}$.
Define the restricted set given by Eq.~(\ref{eq:S}).
Then, construct the proposal state $\{c^\ast_n\}$ as follows.

\textit{Split proposal.}
If $i$ and $j$ belong to the same cluster, i.e., $c_i = c_j$, construct the split proposal $\{c^\mathrm{split}_n\}$ through the following procedure.
\begin{enumerate}
    \item Initialize a \textit{launch} state $\{c^\mathrm{launch}_n\}$.
    Let $c^\mathrm{launch}_i = c_i$ and set $c^\mathrm{launch}_j$ to a new class, e.g., $c^\mathrm{launch}_j = K + 1$. 
    For every $n \in \mathcal{S}$, assign $c^\mathrm{launch}_n$ to either $c_i$ or $K + 1$ with equal probability.
    For every $n \notin \mathcal{S} \cup \{i, j\}$, set $c^\mathrm{launch}_n = c_n$.
    \item Modify the launch state $\{c^\mathrm{launch}_n\}$ through $t$ intermediate restricted Gibbs scans that only update $\{c^\mathrm{launch}_n : n \in \mathcal{S}\}$. In this study, we set $t = 1$.
    In each restricted Gibbs scan, sample $c^\mathrm{launch}_n \in \{c_i, c_j\}$ for every $n \in \mathcal{S}$ according to
    \begin{align}
        ~&~ 
        P(c^\mathrm{launch}_n = k \mid \{c^\mathrm{launch}_m\}_{m \neq n}, \{\bm{\theta}_n\}) \nonumber \\
        =&~ 
        \frac
        {N_k^{\backslash n} p(\bm{\theta}_n \mid c^\mathrm{launch}_n = k, \tau, \{\bm{\theta}_m\}_{m \neq n})}
        {
        N_{c^\mathrm{launch}_i}^{\backslash n} p(\bm{\theta}_n \mid c^\mathrm{launch}_n = c^\mathrm{launch}_i, \tau, \{\bm{\theta}_m\}_{m \neq n}) +
        N_{c^\mathrm{launch}_j}^{\backslash n} p(\bm{\theta}_n \mid c^\mathrm{launch}_n = c^\mathrm{launch}_j, \tau, \{\bm{\theta}_m\}_{m \neq n})
        },
    \end{align}
    where $N_k^{\backslash n} := \sum_{m \neq n} \delta(c_m = k)$.
    Due to the conjugate prior over $\bm{\mu}_k$, $p(\bm{\theta}_n \mid c_n = k, \tau, \{\bm{\theta}_m\}_{m \neq n})$ is analytically evaluated as
    \begin{align}
        ~&~ p(\bm{\theta}_n \mid c_n = k, \tau, \{\bm{\theta}_m\}_{m \neq n}) \nonumber \\
        =&~ 
        \int p(\bm{\theta}_n \mid c_n = k, \bm{\mu}_k, \tau) \, p(\bm{\mu}_k \mid\{\bm{\theta}_m\}_{m \neq n}, \tau) \, \mathrm{d} \bm{\mu}_k \nonumber \\
        =&~ 
        \int \mathcal{N}(\bm{\theta}_n \mid \bm{\mu}_k, \tau^{-1}\mathbf{I}) \, 
        \mathcal{N} \left(
        \bm{\mu}_k
        ~ \middle | ~
        \frac{N_k^{\backslash n}}{N_k^{\backslash n} + \rho} \overline{\bm{\theta}}_k^{\backslash n} + \frac{\rho}{N_k^{\backslash n} + \rho} \bm{\mu}_0, ~
        \frac{1}{\tau(N_k^{\backslash n} + \rho)} \mathbf{I}
        \right) \, \mathrm{d} \bm{\mu}_k \nonumber \\
        =&~
        \mathcal{N} \left(
        \bm{\theta}_n
        ~ \middle| ~
        \frac{N_k^{\backslash n}}{N_k^{\backslash n} + \rho} \overline{\bm{\theta}}_k^{\backslash n} + \frac{\rho}{N_k^{\backslash n} + \rho} \bm{\mu}_0, ~
        \frac{N_k^{\backslash n} + \rho + 1}{\tau(N_k^{\backslash n} + \rho)} \mathbf{I}
        \right).
        \label{eq:marginal_lik}
    \end{align}
    \item Starting from $\{c^\mathrm{launch}_n\}$, perform one additional restricted Gibbs scan to obtain a split proposal $\{c^\mathrm{split}_n\}$.
    \item Compute the (forward) proposal probability $q(\{c^\mathrm{split}_n\} \mid \{c_n\})$,
    which is defined as the transition probability from $\{c^\mathrm{launch}_n\}$ to $\{c^\mathrm{split}_n\}$.
    The reverse proposal probability $q(\{c_n\} \mid \{c^\mathrm{split}_n\})$ is set to one because the reverse move is the deterministic merge of the two labels.
\end{enumerate}

\textit{Merge proposal.}
If $i$ and $j$ belong to the different cluster, i.e., $c_i \neq c_j$, construct the merge proposal $\{c^\mathrm{merge}_n\}$ through the following procedure.
\begin{enumerate}
    \item Initialize a launch state $\{c^\mathrm{launch}_n\}$.
    Let $c^\mathrm{launch}_i = c_i$ and $c^\mathrm{launch}_j = c_j$. 
    For every $n \in \mathcal{S}$, assign $c^\mathrm{launch}_n$ to either $c_i$ or $c_j$ with equal probability.
    For every $n \notin \mathcal{S} \cup \{i, j\}$, set $c^\mathrm{launch}_n = c_n$.
    \item Modify the launch state $\{c^\mathrm{launch}_n\}$ through $t$ intermediate restricted Gibbs scans.
    \item Starting from $\{c^\mathrm{launch}_n\}$, perform one additional restricted Gibbs scan to compute the reverse proposal probability $q(\{c_n\} \mid \{c^\mathrm{merge}_n\})$,
    which is defined as the transition probability from $\{c^\mathrm{launch}_n\}$ to the original state $\{c_n\}$.
    The (forward) proposal probability $q(\{c^\mathrm{merge}_n\} \mid \{c_n\})$ is set to one.
    \item Construct a merge proposal, $\{c^\mathrm{merge}_n\}$. Let $c^\mathrm{merge}_i = c^\mathrm{merge}_j = c_i$. For every $n \in \mathcal{S}$, $c^\mathrm{merge}_n = c_i$. For every $n \notin \mathcal{S} \cup \{i, j\}$, $c^\mathrm{merge}_n = c_n$.
\end{enumerate}

\textit{Metropolis update.}
Finally, we evaluate the proposal state $\{c_n^\ast\}$ (either $\{c^\mathrm{split}_n\}$ or $\{c^\mathrm{merge}_n\}$) with the acceptance probability in Eq.~(\ref{eq:split-merge_ratio_0}).
$p(\{\bm{\theta}_n\} \mid \{c_n\})$ in Eq.~(\ref{eq:split-merge_ratio_0}) denotes the marginal likelihood, which can be decomposed into ``cluster-wise'' likelihoods as,
\begin{align}
    p(\{\bm{\theta}_n\} \mid \{c_n\}, \tau) = \prod_{k=1}^K p(\{\bm{\theta}_n\}_{c_n = k} \mid \tau).
\end{align}
The cluster-wise likelihood $p(\{\bm{\theta}_n\}_{c_n = k} \mid \tau)$ is given by
\begin{align}
    ~&~ p(\{\bm{\theta}\}_{c_n = k} \mid \tau) \nonumber \\
    =&~ \int p(\bm{\mu}_k) \prod_{n=1}^{N} p(\bm{\theta}_n \mid \bm{\mu}_k, \tau)^{\delta(c_n=k)} \, \mathrm{d} \bm{\mu}_k \nonumber \\
    =&~ \int \mathcal{N}(\bm{\mu}_k \mid \bm{\mu}_0, (\rho\tau)^{-1} \mathbf{I}) \,
    \prod_{n=1}^{N} \mathcal{N}(\bm{\theta}_n \mid \bm{\mu}_k, \tau^{-1} \mathbf{I})^{\delta(c_n=k)} \, \mathrm{d} \bm{\mu}_k \nonumber \\
    =&~ \left(\frac{\tau}{2\pi}\right)^{N_k D / 2} \, \left(\frac{\rho}{\rho + N_k} \right)^{D/2}
    \mathrm{exp} \left\{ -\frac{\tau}{2} \left(
    \sum_{n=1}^N \delta(c_n=k) \|\bm{\theta}_n - \overline{\bm{\theta}}_k\|^2 +
    \frac{\rho N_k}{\rho + N_k} \|\overline{\bm{\theta}}_k - \bm{\mu}_0\|^2
    \right) \right\}.
\end{align}

\bibliographystyle{elsarticle-num-names} 
\bibliography{cas-refs.bib}

\end{document}